\documentclass[aps,prd,preprint,groupedaddress,longbibliography,nofootinbib]{revtex4-1}
\usepackage[utf8]{inputenc}
\usepackage{amsmath,amssymb}
\usepackage{amsfonts}
\usepackage{slashed}
\usepackage{xcolor}

\usepackage[pdftex]{hyperref}
\hypersetup{ 
pdfstartview=FitV,
colorlinks=true, 
linkcolor=black, 
citecolor=blue, 
filecolor=black, 
urlcolor=magenta,
}
\usepackage{graphicx}
\usepackage{changes}

\newcommand{\nn}{\nonumber}
\newcommand{\ud}{\mathrm{d}}

\newcommand{\inc}{\textrm{inc}}

\begin{document}

\preprint{CPHT-RR120.122018, NORDITA 2018-116 }

\title{Impact of irrelevant deformations on thermodynamics and transport in
holographic quantum critical states}


\author{Richard A. Davison}
\email{davison@damtp.cam.ac.uk}
%\thanks{}
\affiliation{Department of Applied Mathematics and Theoretical Physics, University of Cambridge, Cambridge CB3 0WA, UK}
\author{Simon A. Gentle}
\email{s.a.gentle@uu.nl}
\affiliation{Institute for Theoretical Physics,  Utrecht University, 3508TD Utrecht, The Netherlands}
\affiliation{Instituut-Lorentz for Theoretical Physics, Leiden University,  2333CA Leiden, The Netherlands}
\author{and Blaise Gout\'eraux}
\email{blaise.gouteraux@polytechique.edu}
\affiliation{Center for Theoretical Physics, \'Ecole Polytechnique, CNRS UMR 7644, Université Paris-Saclay, 91128, Palaiseau, France}
\affiliation{Nordita, KTH Royal Institute of Technology and Stockholm University, Roslagstullsbacken 23, SE-106 91 Stockholm, Sweden}

\date{\today}

\begin{abstract}
We study thermodynamic and transport observables of quantum critical states that arise in the infra-red limit of holographic renormalisation group flows. Although these observables are expected to exhibit quantum critical scaling, there are a number of cases in which their frequency and temperature dependences are in apparent contradiction with scaling theories. We study two different classes of examples, and show in both cases that the apparent breakdown of scaling is a consequence of the dependence of observables on an irrelevant deformation of the quantum critical state. By assigning scaling dimensions to the near-horizon observables, we formulate improved scaling theories that are completely consistent with all explicit holographic results once the dependence on the dangerously irrelevant coupling is properly accounted for. In addition to governing thermodynamic and transport phenomena in these states, we show that the dangerously irrelevant coupling also controls late-time equilibration, which occurs at a rate parametrically slower than the temperature $1/\tau_{eq}\ll T$. At very late times, transport is diffusion-dominated, with a diffusivity that can be written simply in terms of $\tau_{eq}$ and the butterfly velocity, $D\sim v_B^2\tau_{eq}$. We conjecture that in such cases there exists a long-lived, propagating collective mode with velocity $v_s$, and in this case the relation $D=v_s^2\tau_{eq}$ holds exactly in the limit $\tau_{eq} T\gg1$.
\end{abstract}

%\keywords{}

\maketitle

\vskip1cm
\tableofcontents

\section{Introduction} 

\subsection{Context}

Many interesting systems display quantum phase transitions: zero temperature transitions between an ordered and disordered phase when an external parameter (magnetic field, pressure, etc.) is varied. The zero temperature quantum critical point (QCP) at which the transition occurs has a characteristic imprint on the properties of the non-zero temperature state in the `quantum critical wedge' emanating from the QCP \cite{sachdevbook}. For example, the correlation length depends on the temperature $T$ as $\xi\sim T^{-1/z}$ within this wedge, where $z$ is the dynamical critical exponent of the QCP. Other time-independent observables will typically exhibit power law dependencies on temperature, with the powers fixed by the scaling dimensions of the observables. While computing the scaling dimensions often requires intricate higher-loop calculations, once they are known a simple scaling theory can provide a powerful description of the universal physics near the QCP.

However, in practice the usefulness of scaling theories is reduced when observables depend on irrelevant deformations of the QCP. While irrelevant couplings vanish at low energies under RG flow, and so should not affect the physics near the QCP, there are circumstances in which they do (see for instance Chapter 18 of \cite{sachdevbook}). This occurs when the `small' corrections due to the irrelevant coupling are actually the leading contribution to an observable. An intuitive example is the resistivity of a clean QCP without particle-hole symmetry: this vanishes due to translational symmetry and thus `small' corrections to it due to an irrelevant, translational symmetry-breaking deformation of the QCP will strongly depend upon the nature of the irrelevant deformation, see e.g. \cite{Hartnoll:2012rj}. 

In the presence of such dangerously irrelevant deformations, extra information besides the scaling dimension is needed to determine the dependence near the QCP of a given observable on temperature and the irrelevant coupling. In cases where the observable depends on an irrelevant coupling due to its overlap with a slow operator (as in the resistivity example of the previous paragraph), memory matrix methods \cite{forster1975hydrodynamic} provide a powerful way to establish its dependence on temperature and the irrelevant coupling. Outside of these cases, other insights are needed.

Gauge/gravity duality (holography) \cite{Ammon:2015wua,Zaanen:2015oix,Hartnoll:2016apf} is a framework which provides an efficient way to study quantum critical dynamics in a saddle point formulation. By geometrising the RG flow of quantum field theories, such that the energy scale of the field theory is represented by an emergent spatial direction, quantum critical states arise as higher dimensional spacetimes with appropriate scaling symmetries. It is relatively straightforward to construct holographic theories in which such scaling spacetimes arise in the deep interior of spacetimes that are asymptotically AdS i.e.~to construct theories that are conformal in the ultraviolet (UV), and that flow in the infrared (IR) to a variety of quantum critical states \cite{Charmousis:2010zz}. Turning on a small temperature corresponds to introducing a black hole horizon within the deep interior of the spacetime. Observables which are sensitive to the scaling symmetries of the deep interior (or, at non-zero temperatures, to the scaling properties inherited by the horizon) should then exhibit the scaling behaviour characteristic of quantum criticality.

In addition to the dynamical critical exponent $z$, translation-invariant holographic quantum critical states are characterised by two additional exponents $\theta$ and $\Phi$ corresponding to the anomalous dimensions of the entropy and charge densities \cite{Gouteraux:2011ce,Huijse:2011ef,Gouteraux:2013oca, Gouteraux:2014hca,Karch:2014mba,Karch:2015pha,Karch:2015zqd} ($d$ is the spatial dimensionality of the field theory)
\begin{equation}
\label{ScalingHypothesis}
[s]=d-\theta\,,\qquad [\rho]=d-\theta+\Phi\,.
\end{equation}
The authors of \cite{Hartnoll:2015sea} used \eqref{ScalingHypothesis} as a starting point and derived a complete scaling theory of transport observables in quantum critical states of this kind. They showed that a variety of magnetothermoelectric transport observables (but not thermodynamic observables) experimentally measured in cuprate high $T_c$ superconductors could be understood from this phenomenological theory, assuming that these observables are insensitive to irrelevant deformations. The inclusion of non-zero anomalous dimensions is an improvement on previous applications of scaling theories to the cuprates \cite{2005PhRvL..95j7002P} (see \cite{Sachdev:2011cs,2015Natur.518..179K} for more discussion of the role of quantum criticality in the cuprates).

Although the scaling hypothesis \eqref{ScalingHypothesis} was directly inspired by holographic quantum critical states without translational symmetry \cite{Gouteraux:2014hca,Karch:2014mba}, it seemingly fails to describe the properties of an even simpler set of holographic examples \cite{Charmousis:2010zz}: the quantum critical states that arise in the IR of translationally invariant holographic models at non-zero density. To see this, we consider the `incoherent dc conductivity' --  the finite component of the dc conductivity -- whose scaling at low temperatures is governed by the scaling properties of the IR QCP \cite{Davison:2015taa}. This temperature dependence disagrees with that predicted by a scaling theory based on equation \eqref{ScalingHypothesis} \cite{Davison:2015taa}, despite the fact that it doesn't appear to depend on any irrelevant couplings.

The incoherent conductivity quantifies the contribution of diffusive processes (with no momentum drag) to the low frequency conductivity, and is the most important dissipative property of the state. It is of importance even beyond translationally invariant states. Firstly, as it does not overlap with momentum, it should not acquire any dependence on irrelevant deformations which weakly break translational symmetry (unlike other conductivities \cite{Hartnoll:2012rj,Hartnoll:2014gba,Patel:2014jfa}). Secondly, it gives one of the dominant contributions to the dc conductivity of non-Galilean-invariant, pinned charge density wave states \cite{Delacretaz:2017zxd}, which have been experimentally observed both in underdoped and overdoped cuprates and which may persist in the strange metallic region found at optimal doping \cite{Delacretaz:2016ivq}. It has been computed holographically in various translation invariant states in \cite{Hartnoll:2007ip,Jain:2010ip,Chakrabarti:2010xy,Davison:2015taa}, and more recently in states breaking translations in \cite{Amoretti:2017frz,Amoretti:2017axe,Donos:2018kkm,Gouteraux:2018wfe,Amoretti:2018tzw,Andrade:2018gqk}.

\subsection{Summary of results}

In this work, we will ultimately show that not only the incoherent dc conductivity, but also its associated susceptibility and diffusivity (which obeys an Einstein relation), are consistent with a scaling theory based on \eqref{ScalingHypothesis}, but different from that presented in \cite{Davison:2015taa}. At low temperatures, all of these quantities are directly determined by the near-horizon solution, and so we can explicitly determine their dependence on temperature and on irrelevant couplings. While the incoherent dc conductivity always scales with temperature as predicted by the scaling theory, the associated susceptibility and diffusivity do not. However, the temperature scaling of these latter observables is consistent with the scaling theory once their dependence on a dangerously irrelevant coupling is accounted for. We find that such a dangerously irrelevant deformation is present for IR QCPs states with $z=1$ and $\theta\neq0$. We further show that the low frequency scaling of the ac (time-dependent) incoherent conductivity at zero temperature is also consistent with this scaling theory, once its dependence on the irrelevant coupling is accounted for.

One of the main differences between our scaling theory and that of \cite{Davison:2015taa} is that we assign dimensions directly to the incoherent (as opposed to the electrical) conductivity and susceptibility, as these are the near-horizon observables. Our results are further evidence that IR scaling theories are a helpful way to understand the properties of near-horizon observables in holographic theories, but that not all such observables obey naive temperature scaling. In other words, some care must be taken when characterizing the degree to which near-horizon observables are universal.

The dangerously irrelevant deformation also manifests itself by sourcing a long-lived mode at low temperatures in the vicinity of the QCP. This mode has a lifetime $\tau_{eq}$, which schematically takes the form
\begin{equation}
\label{eq:introlonglifetime}
\tau_{eq}\sim \frac1T\left(\frac{T^{\Delta_g}}{g}\right)^2,
\end{equation}
where $g$ is the irrelevant coupling and $\Delta_g$ its dimension. As by definition of an irrelevant deformation $\Delta_g<0$, the timescale $\tau_{eq}\gg1/T$ as $T\to0$. This implies that relativistic hydrodynamics has a significantly restricted range of validity near the QCP, as the dangerously irrelevant coupling slows down the return to equilibrium. This affects the low frequency dependence of the incoherent conductivity, which can display a coherent, Drude-like peak centered at $\omega=0$:
\begin{equation}
\label{eq:sigmaincintro}
\sigma_\inc(\omega) = \frac{\sigma_\inc^{dc}}{1-i\omega\tau_{eq}}.
\end{equation}
A summary of some of our results for translationally invariant systems and their physical implications are given in \cite{Davison:2018ofp}.

We also show that there are analogous results for zero density quantum critical states whose translational symmetry is broken by an irrelevant coupling \footnote{\baselineskip0.5pt These states also have $z=1$ and $\theta\ne0$.}. In particular, these states support a collective excitation with a parametrically long lifetime \eqref{eq:introlonglifetime} set by the symmetry-breaking irrelevant coupling. This long-lived excitation is simply a spatially uniform perturbation of the system's total momentum, and we confirm the result \eqref{eq:introlonglifetime} for the lifetime by an independent memory matrix computation. Furthermore, we show that this excitation leads to a coherent peak in the thermal conductivity (analogously to \eqref{eq:sigmaincintro})
\begin{equation}
\label{eq:kappaacintro}
\bar{\kappa}(\omega) = \frac{\bar{\kappa}_{dc}}{1-i\omega\tau_{eq}},
\end{equation}
and that a consistent scaling theory can account for both the $\omega$ and $T$ dependence of the thermal conductivity, once the dependence on the dangerously irrelevant coupling is properly accounted for.

Irrelevant deformations are also known to play an important role in the thermal diffusivity $D_T$ near the $z=1$ quantum critical points that we study \cite{Blake:2017qgd}. More precisely, for a generic holographic quantum critical state $D_T$, written in units of the Planckian time $\tau_P=\hbar/k_BT$ and the butterfly velocity $v_B$ (an IR velocity that quantifies the spread of quantum chaos \cite{Roberts:2016wdl,Blake:2016wvh}), is a simple universal constant. However, for $z=1$ critical states this universal relation breaks down as $D_T$ becomes sensitive to irrelevant deformations of the critical point. We show that both $D_T$ and $\tau_{eq}$ are in fact controlled by the same dangerously irrelevant coupling and therefore that $\tau_{eq}$ (rather than $\tau_P$) is the timescale controlling thermal diffusion in these systems
\begin{equation}
\label{DTvbtau}
D_T=\frac{2}{d+1-\theta}v_B^2\tau_{eq}\,.
\end{equation}
This is consistent with the conjectures of \cite{Hartman:2017hhp,Lucas:2017ibu}. Near translational invariant quantum critical states, we furthermore show that the usual `incoherent' diffusion constant $D$ of relativistic hydrodynamics is equal to $D_T$ and thus $D\sim v_B^2\tau_{eq}$ in these cases also.

\subsection{Outlook: Diffusivities, IR velocities and timescales}

One of the key results of this work is the identification of the physical timescale $\tau_{eq}$ which controls the thermal diffusivity near $z=1$ IR QCPs through equation \eqref{DTvbtau}. As anticipated in \cite{Hartman:2017hhp,Lucas:2017ibu}, this timescale is the equilibration timescale of the system, which in our case is much longer than that set by temperature $\tau\sim 1/T$, see \eqref{eq:introlonglifetime}. In both cases we have studied (translation-invariant and momentum-relaxing QCPs), this timescale corresponds to the lifetime of the longest-lived non-hydrodynamic excitation near the QCP.
This differs from previous holographic results, where the timescale appearing in \eqref{DTvbtau} was found to be $\tau_{eq}\sim1/T$ \cite{Blake:2016wvh,Blake:2016sud,Blake:2016jnn,Davison:2016auk,Baggioli:2016pia,Kim:2017dgz,Blake:2017ris,Ahn:2017kvc}.

An interesting future direction would be to consider both types of symmetry breaking deformations near the QCP, i.e.~phases at nonzero density and with explicitly broken translations, as in \cite{Gouteraux:2014hca}. Such cases would include examples where there are $z\ne1$ fixed points whose properties are sensitive to dangerously irrelevant deformations. Furthermore we expect that in cases where both deformations are irrelevant near the QCP, the spectrum will display two long-lived modes, capturing the slow relaxation of the incoherent current and of momentum. Accordingly, the frequency dependence of the ac conductivity would be sensitive two both timescales. The interplay between them could pave the way to applications to strange metallic phases, where scenarios with two timescales controlling distinct transport processes have long been advocated for \cite{PhysRevLett.67.2088,PhysRevLett.67.2092}, including in holographic contexts \cite{Blake:2014yla}.

It is natural to ask whether the velocity in the relation \eqref{DTvbtau} should also be interpreted not as the butterfly velocity but as the velocity of a long-lived collective excitation of the system. Indeed, we note that for $z=1$ holographic systems, the speed of sound $v_s^2=2 v_B^2/(d+1-\theta)=1/(d-\theta)$ is precisely equal to the speed in \eqref{DTvbtau}. We anticipate that the systems we study here, in which the $z=1$ spacetime arises at the IR endpoint of the RG flow, will also support a collective mode with this speed and that it is the speed of this mode that sets the diffusivity.

Actually, there is a case where such a mode has already been studied in detail: that of a neutral  fluid with slowly relaxing momentum \cite{Davison:2014lua}. There it was shown that while transport is diffusive at late times, at earlier times the diffusive mode undergoes a collision and turns into a pair of `sound' modes with a long lifetime and speed $v_s$. The thermal diffusivity is $D_T=v_s^2\tau_{eq}$, see equation (2.17) of \cite{Davison:2014lua}. 

Similar relations arise in other hydrodynamic theories where a gapless mode becomes long-lived, for instance in the hydrodynamic theories of fluctuating superconductivity \cite{Davison:2016hno} and fluctuating, pinned charge density waves \cite{Delacretaz:2017zxd}. The velocities are respectively the superfluid sound velocity and the shear transverse sound velocity, respectively. The same relations can also be derived in probe brane scenarios \cite{Karch:2008fa,Davison:2011ek,Chen:2017dsy}, states with generalised global symmetries \cite{Grozdanov:2018ewh}, magnetohydrodynamics and Müller-Israel-Stewart theory \cite{Grozdanov:2018fic}. 

In the more generic cases in which irrelevant deformations are unimportant and therefore $\tau_{eq}$ is not parametrically long, the thermal diffusivity near the QCP is \cite{Blake:2017qgd}
\begin{equation}
D_T=\frac{z}{4\pi(z-1)}\times v_B^2T^{-1}\,.
\end{equation}
In light of the previous discussion, it would be interesting to understand if this relation can also be refined by quantitatively identifying a lifetime $\tau_{eq}=\# T^{-1}$ and velocity $v_s^2=\#v_B^2$ of a collective mode in these systems such that $D_T=v_s^2\tau_{eq}$. Such a relation would indicate that it is not the butterfly velocity $v_B$ that fundamentally sets the thermal diffusivity, but instead the velocity of a collective mode that transports energy through the system. It would also hint at a relation between the velocities of collective modes and the butterfly velocity near quantum critical points.

\section{Holographic QCPs}
\label{section:QCPTI}

We begin by describing the properties of the holographic QCPs of \cite{Charmousis:2010zz,Gouteraux:2011ce,Huijse:2011ef,Gouteraux:2012yr,Gouteraux:2013oca,Donos:2014uba,Gouteraux:2014hca} from a perspective that will prove very useful for understanding the results in the remainder of the paper. A reader who is very familiar with these QCPs can safely skip this section.

The holographic QCPs that we study arise at the IR endpoint of the RG flows generated by relevant deformations of UV conformal fixed points. Using gauge/gravity duality, RG flows of this kind can be captured by the gravitational Einstein-Maxwell-scalar(s) action
\begin{equation}
\label{actionEMD}
S_{UV}=\int\ud^{d+2}x \sqrt{-g}\left(R-\frac12(\partial\phi)^2-\frac{Z(\phi)}{4}F^2-V(\phi)-\frac12 Y(\phi)\sum_{I=1}^d(\partial\psi_I)^2\right).
\end{equation}
The relevant deformations that can be captured by this action include those that break translational symmetry, and those that generate a non-zero density. The classical equations of motion of this action can be found in appendix~\ref{app:eoms}. We are interested in solutions that near the boundary ($u\rightarrow\infty$) have an asymptotic metric that is AdS$_{d+2}$ with unit radius
\begin{equation}
\label{solAdS}
ds^2\rightarrow u^2(-dt^2+d\vec{x}_d^2)+\frac{du^2}{u^2},
\end{equation} 
together with the near-boundary behaviour for the potentials 
\begin{equation}
\label{scalarcouplingsUV}
V(\phi)\rightarrow-d(d+1)+m^2\phi^2/2\,,\qquad Y(\phi)\rightarrow1\,,\qquad Z(\phi)\rightarrow1\,.
\end{equation}

We will always pick an ansatz for the scalar fields $\psi_I$ such that  \cite{Donos:2013eha,Andrade:2013gsa}
\begin{equation}
\psi_I=m \delta_{I j}x^j,
\end{equation}
where $x^j$ are the boundary spatial coordinates.
This means that their equations of motion are automatically satisfied, assuming (as we will throughout this work) that the other fields carry only radial dependence in the background black hole solution. Since $\psi_I$ depend on the spatial coordinates $x^i$, they break translational symmetry. With our choice of UV behaviour $Y(\phi)\rightarrow1$, this breaking is always explicit in the dual field theory. The action for $\psi_I$ is invariant under a shift symmetry $\psi_I\mapsto \psi_I+c_I$, which can be combined with ordinary translations $x^i\to x^i+a^i$ to preserve a diagonal subgroup \cite{Nicolis:2013lma}. This is the underlying reason why the spatial dependence of the $\psi_I$'s is consistent with a radial Ansatz for the background fields. In what follows we won't distiguish between uppercase and lowercase latin indices any longer. 

By choosing the scalar couplings $V(\phi)$, $Y(\phi)$ and $Z(\phi)$ appropriately, we can find gravitational solutions which are dual to field theories that are governed by IR quantum critical points. The metric of these solutions depart from \eqref{solAdS} away from the boundary, and in the deep interior become scale-covariant (at zero temperature). Solutions of this kind arise when the scalar field $\phi$ has a runaway behaviour in the deep interior `IR region', where the scalar couplings can be approximated by exponentials
\begin{equation}
\label{scalarcouplingsIR}
Y(\phi\to\infty)\longrightarrow Y_0 e^{\lambda\phi}\,,\qquad V(\phi\to\infty)\longrightarrow V_0 e^{-\delta\phi}\,,\qquad Z(\phi\to\infty)\longrightarrow Z_0 e^{\gamma\phi}\,,
\end{equation}
where $\lambda$, $\delta$, $\gamma$ are real numbers that will be constrained shortly. Interpreting the radial coordinate as the energy scale of the dual field theory in the usual way, we expect the IR properties of the dual quantum field theory to be governed by this IR region of the spacetime. The effective action governing the IR region is
\begin{equation}
\label{actionEMDIR}
S_{IR}=\int\ud^{d+2}x \sqrt{-g}\left(R-\frac12(\partial\phi)^2-V_0 e^{-\delta\phi}-\frac{Z_0 e^{\gamma\phi}}{4}F^2-\frac12 Y_0 e^{\lambda\phi}\sum_{I=1}^d(\partial\psi_I)^2\right).
\end{equation}
As is now well-understood \cite{Gouteraux:2012yr,Gouteraux:2013oca,Gouteraux:2014hca}, the various terms in the IR effective action are on different footings. The first three terms proportional to the Ricci curvature, the scalar kinetic term and the scalar potential directly support the IR solution dual to the QCP and their effects cannot be neglected \cite{Charmousis:2009xr,Charmousis:2010zz} in order to obtain consistent solutions. The last two terms, proportional to $F^2$ and $\left(\partial\psi\right)^2$, parameterise two deformations of the QCP. Depending on the values of the exponents $\gamma$ and $\lambda$, these deformations will be either marginal or irrelevant. They capture the effects of nonzero density (particle-hole symmetry breaking), and of translation symmetry breaking, near the QCP respectively. When they are marginal, they directly source the IR solution. When they are irrelevant, they source corrections to the IR solution that grow towards the boundary of the IR region of the spacetime. 

We will now describe the IR solutions in greater detail. The action \eqref{actionEMDIR} admits zero temperature scaling solutions \cite{Donos:2014uba,Gouteraux:2014hca}, which are naturally parametrized by two scaling exponents:~$\{z,\theta\}$ \cite{Gouteraux:2011ce,Huijse:2011ef,Gouteraux:2014hca}. These solutions (valid deep in the interior of the bulk) are dual to an IR quantum critical point, and have the form
\begin{equation}
\label{HVsol}
\begin{split}
&\ud s^2=\left(\frac{r}{L}\right)^{2\frac\theta d}\left(-\frac{L^{2z}}{r^{2z}}L_t^2\ud t^2+\frac{\tilde L^2\ud r^2}{r^2}+\frac{L^2}{r^2}L_x^2\ud\vec{x}^2\right),\quad\quad \phi=\kappa\ln\left(\frac{r}L\right)\,,\\
&\kappa^2=\frac2d(d-\theta)(dz-d-\theta)\,,\quad \quad\kappa\delta=2\frac\theta d\,.
\end{split}
\end{equation}
The regime of the full geometry where this IR solution is valid is controlled by the scale $L$: for $\theta<d$ ($\theta>d$), it is valid for $r\gg L$ ($r\ll L$) assuming that that the asymptotically AdS boundary is at $r\rightarrow0$ ($r\rightarrow\infty$). The $r$ coordinate does not extend all the way to the AdS boundary and so is distinct from the radial coordinate $u$ that appears in \eqref{solAdS}. The values of the scales $L_t$ and $L_x$ will be determined in a non-trivial way by the flow to the asymptotically AdS solution \eqref{solAdS}, and so we keep them as free parameters. Their dependence on the UV sources depends on the specific RG flow considered.

Our choice of coordinates makes it manifest that the zero temperature metric transforms covariantly under the scaling
\begin{equation}
\label{eq:IRscaling}
t\to\Lambda^z t\,,\qquad (r,\vec x)\to\Lambda(r,\vec x),
\end{equation}
and therefore that $z$ is the dynamical critical exponent of the critical point. The zero temperature metric is only covariant (rather than invariant) under this transformation when there is a non-zero hyperscaling violation exponent $\theta$. Hyperscaling violation is directly related to the IR running of the scalar $\phi$, and to the fact that the scale $L$ has not decoupled from the IR theory. $\theta$ determines the effective spatial dimensionality $d-\theta$ of the IR quantum critical state. This statement can be made precise by embedding such solutions in higher or lower-dimensional spacetimes \cite{Kanitscheider:2009as,Gouteraux:2011ce,Gouteraux:2011qh,Faedo:2016cih}.

We have not yet given an expression for the value of the dynamical exponent $z$. To do so, we need to consider the deformations of the QCP due to the gauge field $F$ and scalar field $\psi$ terms in the IR effective action \eqref{actionEMDIR}. For simplicity we will turn on one or the other but not both together. That is, we consider either translation-invariant, nonzero density states, or translation-breaking, zero density states.

\subsection{Marginal deformation $(z\ne1)$}

The first possibility is that the deformation of interest is marginal. That is, it does not give rise to terms with different powers of $r$ in the solution \eqref{HVsol}. For the nonzero density cases, this implies
\begin{equation}
\label{CouplingsIRzneq1}
\begin{split}
A=A_0\left(\frac{r}L\right)^{\theta-d-z}\,L_t\ud t\,,\quad&\quad A_0^2=\frac{2(z-1)}{Z_0(d+z-\theta)}\,,\\ \kappa\gamma=2d -2(d-1)\frac\theta d\,,\quad &\quad  \tilde L^2=\frac{(d+z-\theta)(d-1+z-\theta)}{-V_0}\,,
\end{split}
\end{equation}
while for the translation-breaking cases, it implies
\begin{equation}
\label{CouplingsIRzneq1m}
\kappa\lambda=-2\,,\qquad L_x^2=\frac{(dz-\theta)m^2}{-2(z-1)V_0}\,,\qquad  \tilde L^2=\frac{(dz-\theta)(d+z-\theta)}{-V_0}\,.
\end{equation}
In both cases, the IR metric enjoys nonrelativistic scaling $z\neq1$. The relations between $\left(\delta,\gamma,\lambda\right)$ and $(z,\theta)$ can be inverted to express $z$ and $\theta$ in terms of parameters in the effective action. Thus the scaling exponents of the IR QCP are completely determined by specifying the gravitational action.

When a small temperature $T$ is turned on, the solutions with marginal deformations change: there is a horizon at $r=r_h$ and
\begin{equation}
\label{HVsolT}
\ud s^2=\left(\frac{r}{L}\right)^{2\frac\theta d}\left(-\frac{L^{2z}f}{r^{2z}}L_t^2\ud t^2+\frac{\tilde L^2\ud r^2}{r^2f}+\frac{L^2}{r^2}L_x^2\ud\vec{x}^2\right),\quad f=1-\left(\frac{r}{r_h}\right)^{d+z-\theta}\,.
\end{equation}
These IR solutions are dual to the thermal state of the quantum critical theory, with $T$ related to $r_h$ by \eqref{eq:tempdimension} below.

Even at non-zero temperature, there is a region $L\ll r\ll r_h$ where the metric looks like the zero temperature form \eqref{HVsol}. One simple example of this is the AdS-Reissner-Nordström solution, which is characterized by an AdS$_2\times$R$^2$ zero temperature IR geometry. The AdS$_2$ can be placed at nonzero temperature, which describes small departures from the zero temperature state.

Imposing the null energy condition and positivity of the low temperature heat capacity results in the following restrictions on the allowed parameter space of IR solutions
\begin{equation}
\label{eq:EMDparameterspace}
\frac{d-\theta}{z}\geq0\,,\quad (d-\theta)(dz-d-\theta)\geq0\,,\quad (z-1)(d+z-\theta)\geq0\,.
\end{equation}

\subsection{Irrelevant deformation $(z=1)$}
\label{sec:irrdefsubsec}

The second possibility we would like to consider is that the deformation of interest is irrelevant. For these cases the $T=0$ IR solution has $z=1$, and the deformation sources power law (in $r$) corrections to this solution that grow towards the edge of the IR region.
This means that the IR solution is like a `CFT' in $d-\theta$ spatial dimensions, in the presence of an irrelevant deformation parameterised by a coupling $g=\left\{A_0,m\right\}$. As dilatations are broken by $\theta\neq0$, it is not an actual CFT, but it can be endowed with a generalized conformal structure in the sense of \cite{Kanitscheider:2009as,Gouteraux:2011qh}. The full solution is obtained by solving for the backreaction of the irrelevant field(s) $\left\{A,\psi_i\right\}$ order-by-order in $g$. The leading corrections take the form (where the subscript $g=0$ means \eqref{HVsol} with $z=1$)
\begin{equation}
\label{IrrDefz=1}
\ud s^2=\ud s^2_{g=0}\left(1+\# g^2 \left(\frac{r}L\right)^{2\Delta_g}+O(r^{4\Delta_g})\right),\quad \phi=\phi_{g=0}+\# g^2 \left(\frac{r}L\right)^{2\Delta_g}+O(r^{4\Delta_g})\,.
\end{equation}
The corrections are quadratic in the irrelevant coupling as the corresponding fields appear quadratically in the scalar and Einstein equations.

For non-zero density cases, \eqref{IrrDefz=1} is supplemented by
\begin{equation}
\label{CouplingsIRzeq1}
A_t{}=L_tA_0 \left(\frac{r}L\right)^{\theta-d-1-2\Delta_g}\left(1+\# A_0^2 \left(\frac{r}L\right)^{2\Delta_{A_0}}+O(r^{4\Delta_{A_0}})\right)\,,\quad 2\Delta_{A_0}=2(d-\theta)-\kappa\gamma+\frac2d\theta\,.
\end{equation}

For the translational symmetry breaking cases, \eqref{IrrDefz=1} is supplemented by
\begin{equation}
\label{CouplingsIRzeq1m}
2\Delta_m=2+\kappa\lambda\,.
\end{equation}

The critical states \eqref{HVsol} with $z=1$ can also be placed at a small, nonzero temperature by introducing an event horizon at $r=r_h$. Close to the horizon, the metric will have the form \eqref{HVsolT} with $z=1$. The zero temperature form \eqref{IrrDefz=1} will be recovered in the range $L\ll r\ll r_h$. 

Since $z=1$, the inequalities \eqref{eq:EMDparameterspace} enforce that $\theta<0$. This in turn implies that the expansions in \eqref{IrrDefz=1} can only make sense if $\Delta_g<0$ (the IR is located at $r\to+\infty$ in these coordinates). This is precisely what we expect for the dimension of an irrelevant coupling. The irrelevant coupling $g$ breaks the $z=1$ symmetry of the QCP, in tandem with breaking either the particle-hole symmetry or the translational symmetry of the QCP. $g$ also breaks Lorentz boost symmetry in both cases. The dimension of the irrelevant coupling is primarily determined by the value of $\gamma$ or $\lambda$, respectively.

\subsection{More on irrelevant deformations of holographic QCPs}

We will now characterise more precisely the scaling properties of the $z=1$ IR solutions by relating $\Delta_{A_0}$ and $\Delta_m$ (defined above) to the dimensions of irrelevant couplings. We will begin with the case where the irrelevant deformation breaks particle-hole symmetry ($A_t\ne0$). $A_t$ has two independent modes: in addition to the $A_0$ mode that scales as $r^{\theta-d-1+2\Delta_{A_{0}}}$, there is a constant $r^0$ mode that is allowed by U(1) gauge invariance. As usual in holographic theories, we would like to interpret one of these modes as the source of an operator that is irrelevant near the IR critical point, and the other as the corresponding expectation value \cite{Ammon:2015wua,Zaanen:2015oix,Hartnoll:2016apf}. From their radial dependence, we see that the difference between the IR scaling dimension of the source and the expectation value is $\theta-d-1+2\Delta_{A_{0}}$. Combining this with the fact that their dimensions should sum to $d+1-\theta$, due to the effective dimensionality of the IR critical point, we are left with two possible values for the dimensionality of the operator. We choose the positive value $\Delta_{irr}=d+1-\theta-\Delta_{A_{0}}>0$, and so the source has dimension $\Delta_{A_{0}}<0$. Note that since $\Delta_{irr}>d+1-\theta$, this is an irrelevant deformation of the IR critical point. 

We interpret the mode which grows fastest near the boundary of the IR solution as the source, and hence $A_0$ is the source for an irrelevant deformation of the IR critical point with dimension $\Delta_{A_{0}}$ given by \eqref{CouplingsIRzeq1}.
This irrelevant source produces a deformation of the metric which vanishes in the deep IR, as expected. This is in contrast to the constant mode (the expectation value), which would backreact on the metric in a way that grows in the IR and destroys the critical point, consistent with the discussion in \cite{Kiritsis:2016kog}.

For the case where the irrelevant operator breaks translational symmetry, we can similarly identify $m$ with the source of an irrelevant operator in the IR, with the following scaling dimensions for $m$ and for the corresponding IR operator (obtained by a similar argument to above)
\begin{equation}
\label{DimPsis}
\Delta_m=1+\frac{\kappa\lambda}2<0\,,\quad \Delta_{irr}=d+1-\theta-(\Delta_m-1)>d+2-\theta\,.
\end{equation}
There is a shift in the dimension of $\Delta_{irr}$ (i.e.~it does not obey the naive equality $\Delta_m+\Delta_{irr}=d+1-\theta$) due to the spatial dependence of the source at the IR boundary. This is analogous to the Harris criterion for randomly disordered sources \cite{0022-3719-7-9-009}. 

From \eqref{CouplingsIRzneq1}, \eqref{CouplingsIRzneq1m}, we see that for the $z\neq1$ QCPs, $g=\left\{A_0,m\right\}$ is a marginal coupling.

\subsection{IR scaling of thermodynamic observables \label{subsection:IRscalingThermo}}

Physically, we expect that the scaling dimension of the irrelevant coupling will control the IR behaviour of certain observables. In this section, we specifically comment on thermodynamic observables. The remainder of the paper will in large part be devoted to the study of the impact of these irrelevant deformations on transport.

The scaling properties of the solutions we have just described result in many field theory observables exhibiting scaling behaviour at low temperatures and frequencies. This scaling behaviour can be understood in terms of the anomalous IR scaling dimensions of entropy and charge density in these solutions \cite{Gouteraux:2011ce, Huijse:2011ef, Gouteraux:2013oca,Gouteraux:2014hca,Karch:2014mba}.

Once the values of $V_0$ and $Z_0$ in the IR action have been fixed, the zero temperature IR solution is characterized by the parameters $L_t,L_x$ and $L$. Changing the values of these parameters does not induce any RG flow (i.e.~any new radial dependence in the IR solution) and so they are marginal. There are two important deformations that do change the radial dependence of the IR solution. The first is turning on $r_h$ i.e.~turning on a non-zero temperature. The second is turning on the coupling $g=\left\{A_0,m\right\}$, which for $z=1$ solutions is an irrelevant deformation that induces an RG flow \footnote{\baselineskip0.5pt There are additional irrelevant deformations, but since they do not affect the low temperature and low frequency behaviour of the observables of interest in this work, we will not discuss them.}. 

We will now assign IR scaling dimensions by determining how quantities depend on these two IR scales -- the temperature $T$ and the irrelevant coupling $g$. This is straightforward for quantities which can be expressed explicitly in terms of the near-horizon gravitational solution, as their $T$ and $g$ dependence is then manifest. The Hawking temperature is related to the horizon radius by
\begin{equation}
\label{eq:tempdimension}
T=\frac{(d+z-\theta)}{4\pi\tilde L}L_t \left(\frac{r_h}{L}\right)^{-z}\quad\Rightarrow\quad [T]=z,
\end{equation}
in our conventions where $[r]=-1$. This result is consistent with interpreting $T$ as an inverse time, where $[t]=-z$ in line with the scaling transformation (\ref{eq:IRscaling}). The entropy density $s$ and charge density $\rho$ (for the nonzero density states) can be calculated from the area of the horizon and the electric flux emitted by the horizon, and are given by
\begin{equation}
\label{eq:holothermo}
s=4\pi L_x^d \left(\frac{r_h}{L}\right)^{\theta-d},\quad\quad\quad \rho=L_x^d\frac{(\theta-d-z-2\Delta_g)Z_0}{\tilde L}A_0.
\end{equation}
 We can therefore assign them the dimensions
\begin{equation}
[s]=d-\theta,\quad\quad\quad\quad\quad [\rho]=\Delta_{A_0},
\end{equation}
using equation (\ref{eq:tempdimension}). 

It is also convenient to define an anomalous scaling dimension $\Phi$ for the charge density \cite{Karch:2014mba,Hartnoll:2015sea} via \footnote{\baselineskip0.5pt Note that here we have assigned a different value to $\Phi$ for $z\neq1$ solutions than in \cite{Davison:2015taa}.}
\begin{equation}
\label{RelationPhiZeta}
[\rho]=\Delta_{A_0}=d-\theta+\Phi.
\end{equation}
We observe that even for the marginal case, $\Phi=\theta-d$ is non trivial and implies $[\rho]=0$. We will see later that this scaling assignment correctly reproduces the explicit low $T$ or low frequency scaling of transport observables.

We emphasise that these scaling dimensions do not tell us anything about how $\rho$ and $s$ depend on the sources which deform the UV CFT, such as the chemical potential $\mu$. At zero temperature, the UV sources generate an RG flow to the IR solution, where the values of all of the parameters in the IR solution $A_0,L_t,L_x$ etc.~will depend non-trivially on the UV sources. Access to the entire RG flow is needed to reconstruct this dependence. But the dependence of $\rho$ and $s$ on $r_h$ does tell us their $T$ dependence at low $T$, as the other parameters in the IR solution $A_0,L_t,L_x$ etc.~are $T$-independent in the limit $T\rightarrow0$ (otherwise the zero temperature IR solution would not exist).

For the $z\ne1$ solutions, $[A_0]=0$ (i.e.~this coupling is marginal) and so the scaling dimensions of $s$ and $\rho$ indicate their dependence on the only dimensionful scale $T$ \footnote{\baselineskip0.5pt In fact, as $A_0$ can be replaced with $Z_0$ by equation \eqref{CouplingsIRzneq1} for these solutions, it is not very meaningful to ask how IR observables depend on $A_0$.}, $s\sim T^{(d-\theta)/z}$ and $\rho\sim T^0$. For $z=1$ solutions, the total scaling dimensions of $s$ and $\rho$ tell us the combined dependence on $T$ and the dimensionful irrelevant coupling $A_0$. The information about how they separately depend on $A_0$ and on $T$ is not captured by the total dimension. More information -- the explicit expressions \eqref{eq:holothermo} -- are needed to determine this separate dependence, $s\sim (A_0)^0\, T^{d-\theta}$ and $\rho\sim A_0\, T^0$.

As we have emphasised, in general the relation between the UV sources and the IR sources of the theory is not simple. However, the fact that the charge density $\rho$ \eqref{eq:holothermo} is directly proportional to $A_0$ at zero temperature means that it can sometimes be helpful to think of $A_0$ as an `IR charge density'.

\section{Translation invariant case}

We expect the IR solutions described above to control the low energy properties of the field theory states dual to the complete, asymptotically AdS gravitational solutions. We are now going to explore a variety of dynamical response properties of the field theory states and explain how their dependence on temperature $T$ and frequency $\omega$ (at low $T$ and $\omega$) can be understood in terms of the IR QCPs we have just described. We will focus on the linear response dynamics of charge density and energy density, which are responsible for the transport properties of the system. In this section, we will study cases with translational symmetry but non-zero density $\left(m=0,A_t\ne0\right)$.

\subsection{Incoherent diffusion in linearised hydrodynamics \label{section:thermohydro}}

We begin by describing the theory of hydrodynamics that we expect to govern the transport properties of our field theory over sufficiently long timescales and distances. We will mainly focus on this theory's `incoherent' diffusive mode, which characterises transport in the absence of momentum flow. In later subsections we will show how the diffusivity of this mode, as well as the corresponding conductivity and susceptibility, are governed at low $T$ and $\omega$ by the IR QCP. This is in contrast to processes involving momentum flow, which are sensitive to the UV translational symmetry.

\subsubsection{Excitations in linearised hydrodynamics \label{section:linearhydro}}

Linearised hydrodynamics is the theory that describes the transport of small perturbations of charge and heat over long distance and timescales in a system which is locally in thermal equilibrium. We will assume that our system has both translational and rotational symmetry, but no particular form of boost symmetry \cite{Hartnoll:2016apf},\footnote{\baselineskip0.5pt A related analysis in a different frame has recently appeared in \cite{deBoer:2017abi}.}. 

The hydrodynamic variables are the long wavelength perturbations of the entropy density $\delta s$, charge density $\delta\rho$ and momentum density $\pi^i$. In the absence of external sources, the linearized conservation equations are
\begin{equation}
\begin{aligned}
\label{eq:conservationeqs}
&\partial_t \delta s+\partial\cdot \left(\delta q/T\right)=0,\\
&\partial_t\delta \rho+\partial\cdot \delta j=0,\\
&\partial_t \delta\pi^i+\partial_j \delta\tau^{ji}=0.
\end{aligned}
\end{equation}
The perturbations $(\delta s,\delta\rho,\delta\pi^i)$ are sourced by temperature, chemical potential, and velocity perturbations $(\delta T,\delta\mu,\delta v^i)$. The static susceptibility matrix $\chi$ relating these quantities has the form
\begin{equation}
\left(\begin{array}{c} \delta\rho \\ \delta s \\ \delta\pi^i\end{array}\right)=\left(
\begin{array}{ccc}
\chi_{\rho\rho}&\chi_{\rho s}&0\\
\chi_{\rho s}&\chi_{ss}&0\\
0&0&\chi_{\pi\pi}
\end{array}
\right)\left(\begin{array}{c} \delta \mu\\ \delta T \\ \delta v^i\end{array}\right).
\end{equation}
In $d$ spatial dimensions, and in the absence of external sources, the constitutive relations for the charge, heat, and momentum currents are (neglecting terms of order $\partial^2$ and higher)
\begin{align}
\label{JconstRel}
&\delta j^i=\rho \delta v^i-\sigma_0\partial^i\delta\mu-\alpha_0\partial^i\delta T,\\
\label{QconstRel} &\delta q^i=s T \delta v^i-T\alpha_0\partial^i \delta\mu-\bar\kappa_0\partial^i\delta T,\\
\label{TauconstRel} &\delta \tau^{ij}=\delta^{ij}\delta p-\eta\left(\partial^i \delta v^j+\partial^j \delta v^i-\frac{2}{d}\delta^{ij}\left(\partial\cdot \delta v\right)\right) -\zeta\left(\partial\cdot \delta v\right)\delta^{ij},
\end{align}
where $\delta p = s\delta T + \rho\delta\mu$ is the pressure fluctuation, $\eta$ and $\zeta$ are the shear and bulk viscosities and $\sigma_0,\alpha_0$ and $\bar{\kappa}_0$ are further dissipative transport coefficients related to the thermoelectric conductivities $\sigma,\alpha,\bar{\kappa}$ by 
\begin{equation}
\label{hydrocond}
\sigma(\omega)=\frac{\rho^2}{\chi_{\pi\pi}}\frac{i}{\omega}+\sigma_0\,,\quad \alpha(\omega)=\frac{s\rho}{\chi_{\pi\pi}}\frac{i}{\omega}+\alpha_0\,, \quad \bar\kappa(\omega)=\frac{Ts^2}{\chi_{\pi\pi}}\frac{i}{\omega}+\bar\kappa_0.
\end{equation}
The divergent low frequency parts of the thermoelectric conductivities are a consequence of the non-zero static susceptibilities between the thermoelectric currents and the conserved momentum \cite{Hartnoll:2012rj}
\begin{equation}
\label{OffDiagMomSusceptibilities}
\chi_{\pi j}=\rho\,,\quad\chi_{\pi q}=s T.
\end{equation}

The longitudinal excitations of this theory consist of a diffusive excitation with dispersion relation
\begin{equation}
\label{eq:diffdispersion}
\omega_D=-iDk^2,\quad\quad D=\frac{s^2T^2\sigma_0+\bar{\kappa}_0T\rho^2-2\rho sT^2\alpha_0}{T^2\left(s^2\chi_{\rho\rho}+\rho^2\chi_{ss}-2s\rho\chi_{\rho s}\right)},
\end{equation}
as well as two sound modes with dispersion relations
\begin{equation}
\begin{aligned}
\label{eq:sounddispersion}
&\omega^\pm=\pm\sqrt{\frac{\rho^2\chi_{ss}+s^2\chi_{\rho\rho}-2s\rho\chi_{s\rho}}{\chi_{\pi\pi}\left(\chi_{ss}\chi_{\rho\rho}-\chi_{s\rho}^2\right)}}k- \frac{i\Gamma}{2}k^2,\\
&\Gamma=\frac{2\eta\left(1-\frac{1}{d}\right)+\zeta}{\chi_{\pi\pi}}-D+\frac{\sigma_0\chi_{ss}-2\alpha_0\chi_{\rho s}+\frac{\bar{\kappa}_0}{T}\chi_{\rho\rho}}{\chi_{ss}\chi_{\rho\rho}-\chi_{s\rho}^2}.
\end{aligned}
\end{equation}
The sound waves are `coherent' \cite{Hartnoll:2014lpa} excitations that transport perturbations of both long-lived momentum density $\delta\pi ^i$ and of pressure $\delta p$ through the system. The importance of pressure fluctuations in the transport of momentum density is obvious from the form of the stress tensor (\ref{TauconstRel}).

In contrast to this, the diffusive mode which is the focus of this paper is an `incoherent' excitation of the system, in that it does not transport long-lived momentum density. We can define perturbations of an `incoherent' density $\delta\rho_{\text{inc}}$ \footnote{\baselineskip0.5pt We have adopted a slightly different normalization than in \cite{Davison:2015taa}, which proves mode convenient as it removes some dependence on microscopic parameters in our later analysis.}
\begin{equation}
\label{RhoInc}
\delta\rho_{\text{inc}}\equiv s^2T\delta\left(\rho/s\right)=T\left(s\delta\rho-\rho\delta s\right),
\end{equation}
which obeys the conservation equation
\begin{equation}
\label{IncConsEq}
\partial_t\delta\rho_{\text{inc}}+\partial\cdot \delta j_{\text{inc}}=0,\quad \quad \delta j_{\text{inc}}\equiv sT\delta j-\rho \delta q.
\end{equation}
The incoherent perturbations $\delta\rho_{\text{inc}}$ and $\delta j_{\text{inc}}$ are special because they do not overlap with fluctuations in the pressure and momentum density i.e.~their static susceptibilities vanish
\begin{equation}
\chi_{\rho_{\text{inc}}p}=0,\quad\quad\quad\chi_{j_{\text{inc}}\pi}=0.
\end{equation}
The consequence of this is, that to leading order in the hydrodynamic limit \footnote{\baselineskip0.5pt Explicitly, we let $\omega\to\lambda^2\omega$, $k\to\lambda k$ and expand the Green's function at leading order in $\lambda$.}  the retarded Green's function of $\delta\rho_\text{inc}$ has the simple diffusive form
\begin{equation}
\label{DiffusiveGRrhoInc}
G^R_{\rho_{\text{inc}}\rho_{\text{inc}}}(\omega,k)=\frac{-k^2\left(T^2 s^2\sigma_0-2\rho s T^2\alpha_0+\rho^2T\bar\kappa_0\right)}{-i\omega+Dk^2}.
\end{equation}
It does not have poles corresponding to the propagation of the sound waves (\ref{eq:sounddispersion}) \footnote{\baselineskip0.5pt i.e.~the full Green's function does not have sound poles when expanded in the limit $\lambda\rightarrow0$ with $\omega\rightarrow\lambda \omega$, $k\rightarrow\lambda k$.}. The diffusivity $D$ in equation (\ref{eq:diffdispersion}) obeys the Einstein relation
\begin{equation}
\label{EinsteinInc}
D=\frac{\sigma_{\text{inc}}}{\chi_{\text{inc}}},
\end{equation}
where $\sigma_{\text{inc}}$ and $\chi_{\text{inc}}$ are the dc conductivity and static susceptibility of $\delta\rho_{\text{inc}}$
\begin{equation}
\begin{aligned}
\label{eq:incoherentobjectsdefn}
&\sigma_{\text{inc}}\equiv\lim_{\omega\rightarrow0}\lim_{k\rightarrow0}\frac{i}{\omega}G^R_{j_{\text{inc}}j_{\text{inc}}}(\omega,k)=\lim_{\omega\rightarrow0}\lim_{k\rightarrow0}\frac{i}{\omega}\frac{\omega^2}{k^2}G^R_{\rho_{\text{inc}}\rho_{\text{inc}}}(\omega,k)=T^2 s^2\sigma_0-2\rho s T^2\alpha_0+\rho^2T\bar\kappa_0,\\
&\chi_{\text{inc}}\equiv-\lim_{k\rightarrow0}\lim_{\omega\rightarrow0}G^R_{\rho_{\text{inc}}\rho_{\text{inc}}}(\omega,k)=T^2\left(\rho^2\chi_{ss}-2s\rho\chi_{\rho s}+s^2\chi_{\rho\rho}\right).
\end{aligned}
\end{equation}
Note that in contrast to the dc conductivities of charge and heat individually, the dc conductivity of $\delta\rho_{\text{inc}}$ is finite. This is because $\delta j_{\text{inc}}$ has no overlap with the conserved momentum \cite{Davison:2015taa}. It is this independence from momentum conservation that makes it possible for $\sigma_{\text{inc}}$ to be controlled by the underlying IR QCP (as we will show in the next subsection), unlike the electrical conductivity. Also note that while the values of the incoherent conductivity and susceptibility depend on the overall normalization of $\delta\rho_{\text{inc}}$, the value of the diffusivity $D$ does not.

\subsubsection{The incoherent susceptibility}

We have identified a particular linear combination of $\delta\rho$ and $\delta s$ (that given by $\delta\rho_{\text{inc}}$) as the object that diffuses. It is convenient to change variables from $\left(\delta\rho,\delta s\right)$ to the pair $\left(\delta\rho_{\text{inc}},\delta p\right)$ which are orthogonal in the sense that $\chi_{\rho_{\text{inc}}p}=0$. 
The sources for these variables are
\begin{equation}
\begin{aligned}
\label{HydroSources}
&\delta s_{\text{inc}}=\frac{1}{T\left(\rho^2\chi_{ss}+s^2\chi_{\rho\rho}-2s\rho\chi_{s\rho}\right)}\left(\left(s\chi_{\rho s}-\rho\chi_{ss}\right)\delta T + \left(s\chi_{\rho\rho}-\rho\chi_{\rho s}\right)\delta\mu\right),\\
&\delta s_p=\frac{\chi_{ss}\chi_{\rho\rho}-\chi_{s\rho}^2}{\rho^2\chi_{ss}+s^2\chi_{\rho\rho}-2s\rho\chi_{s\rho}}\left(s\delta T+\rho\delta\mu\right),
\end{aligned}
\end{equation}
respectively. The source for the incoherent current $j_{\text{inc}}$ is $E_{\text{inc}}=-\partial\delta s_{\text{inc}}$. In the $(\delta\rho_{\text{inc}},\delta p)$ basis, the susceptibility matrix diagonalizes and its second entry reads
\begin{equation}
\chi_{pp}\equiv\left.\frac{\delta p}{\delta s_p}\right|_{\delta s_{\textrm{inc}}=0}=\frac{\left(\rho^2\chi_{ss}-2s\rho\chi_{\rho s}+s^2\chi_{\rho\rho}\right)}{\left(\chi_{\rho\rho}\chi_{ss}-\chi_{\rho s}^2\right)}\,.
\end{equation}
By turning on the sources $\delta T$ and $\delta\mu$ such that $\delta s_p=0$, only perturbations in the incoherent density $\delta\rho_{\text{inc}}$ will be sourced. This allows us to write the incoherent susceptibility as
\begin{equation}
\label{chiIncdeltasp0}
\chi_{\text{inc}}=\left(\frac{\delta\rho_\text{inc}}{\delta s_\text{inc}}\right)_{\delta s_p=0}=sT\left(\frac{\delta\rho_\text{inc}}{\delta\mu}\right)_{\delta s_p=0}=-\rho T\left(\frac{\delta\rho_\text{inc}}{\delta T}\right)_{\delta s_p=0}.
\end{equation}

Another useful expression for $\chi_{\text{inc}}$ can be found by using thermodynamic identities on the expression in equation (\ref{eq:incoherentobjectsdefn})
\begin{equation}
\label{ChiIncSusc}
\begin{aligned}
\chi_{\text{inc}}&=T^2\left(\rho^2\chi_{ss}-2s\rho\chi_{\rho s}+s^2\chi_{\rho\rho}\right)\\
&=T^2\rho^2\left(\frac{\partial s}{\partial T}\right)_\mu+T^2s\left(s\left(\frac{\partial\rho}{\partial\mu}\right)_T-2\rho\left(\frac{\partial s}{\partial\mu}\right)_T\right)\\
&=T\rho^2 c_\rho+T^2\left(\frac{\partial\rho}{\partial\mu}\right)_T\left(s-\rho\left(\frac{\partial s}{\partial\rho}\right)_T\right)^2,
\end{aligned}
\end{equation}
where $c_\rho=T\left(\partial s/\partial T\right)_\rho$ is the specific heat at fixed charge density.
This indicates that, in general, we need to know numerous thermodynamic properties of the system to determine $\chi_{\text{inc}}$. However, in certain limits $\chi_{\text{inc}}$ is dominated by one of the terms in the expression above. The limit of interest to us is that of a non-zero density state at low temperature. We assume that, in this low temperature limit, the charge susceptibility is finite while the entropy density is a power law $s\sim T^\alpha$ with $\alpha\geq0$. These conditions will be valid in all of the holographic theories we examine later. For $\alpha\ne0$, the dominant term at low temperatures is then
\begin{equation}
\label{LowTIncSusc}
\chi_{\text{inc}}\left(T\rightarrow0\right)\longrightarrow T\rho^2c_\rho,
\end{equation}
and so the incoherent susceptibility is simply proportional to the heat capacity at constant charge density. It is this identity which will allow us to show that $\chi_{\text{inc}}$ is governed by the IR QCP in the holographic theories described in the previous section. The case $\alpha=0$ is more subtle, but for the type of holographic theories we are ultimately concerned with the results of \cite{Blake:2016jnn} can be used to show that \eqref{LowTIncSusc} is valid even in this case.

We can also evaluate the susceptibility of pressure at low temperatures to find
\begin{equation}
\chi_{pp}(T\rightarrow0)\rightarrow \frac{\rho^2}{\chi_{\rho\rho}}\,.
\end{equation}
In holographic theories at non-zero density, the chemical potential $\mu$ of the theory is not a near-horizon property, but depends on knowledge of the entire spacetime. As a consequence, the low temperature static charge susceptibility $\chi_{\rho\rho}=\left(\partial\rho/\partial\mu\right)_T$ at low temperature is not an IR property of these theories.

By diagonalizing the susceptibility matrix in the basis $(\delta\rho_{\text{inc}},\delta p)$, we have separated its components into IR and UV-dominated pieces.

\subsubsection{Linearised hydrodynamics in an electric field \label{subsection:hydrowithE}}

The incoherent conductivity also controls the dissipative dynamics of another set of physical processes in this system. 

The hydrodynamic equations (\ref{eq:conservationeqs}) to (\ref{TauconstRel}) are valid in the absence of external sources. In the presence of a small electric field $\delta E_i$, the momentum conservation equation should be modified to
\begin{equation}
\partial_t \delta\pi^i+\partial_j \delta\tau^{ji}=\rho\delta E_i,
\end{equation}
while $\partial_i\delta\mu$ should be replaced with $\partial_i\delta\mu-\delta E_i$ in the constitutive relations (\ref{JconstRel}) to (\ref{TauconstRel}). The presence of an external electric field affects the measured conductivities and diffusivities. One experimentally relevant configuration is when there is an electric field such that no current flows (open-circuit boundary conditions). The open circuit dc thermal conductivity is simply related to the incoherent conductivity by
\begin{equation}
\label{eq:kappahydro}
\kappa\equiv-\frac{\delta j_s}{\partial \delta T}\biggr|_{\delta j=0}=\frac{\sigma_\text{inc}}{T\rho^2}.
\end{equation}
Similarly, under the condition $\partial\cdot \delta j=0$ (i.e.~charge perturbations are static $\partial_t\delta\rho=0$), there is a hydrodynamic diffusion equation
\begin{equation}
\partial_t\left(\delta T +\frac{\chi_{\pi\pi}\left(\rho\alpha_0-s\sigma_0\right)}{\rho^2\left(\partial s/\partial T\right)_\rho}\partial\cdot\delta v\right)=D_T\partial^2\left(\delta T +\frac{\chi_{\pi\pi}\left(\rho\alpha_0-s\sigma_0\right)}{\rho^2\left(\partial s/\partial T\right)_\rho}\partial\cdot\delta v\right)+O(\partial^3)
\end{equation}
with a `thermal' diffusivity
\begin{equation}
\label{eq:thermaleq}
D_T\equiv\frac{\kappa}{c_\rho},
\end{equation}
which is simply related to $\sigma_{\text{inc}}$ by equation (\ref{eq:kappahydro}).

In the low temperature limit described above, $D$ therefore coincides with $D_T$
\begin{equation}
\label{eq:relationbetweenDs}
D_T=\frac{\kappa}{c_\rho}=\frac{\sigma_{\text{inc}}}{T\rho^2c_\rho}=\lim_{T\rightarrow0}\frac{\sigma_{\text{inc}}}{\chi_{\text{inc}}}=\lim_{T\rightarrow0}D,
\end{equation}
due to the relation (\ref{LowTIncSusc}).

\subsubsection{Lorentz and conformally invariant systems}

A special case of the previous discussion are systems with microscopic Lorentz symmetry, which is the case for all of the holographic theories we will discuss in the next section. In the Lorentz case (which applies to asymptotically AdS holographic theories amongst others), this symmetry results in the relations (e.g. \cite{Kovtun:2012rj})
\begin{equation}
\label{RelativisticLimit}
\bar\kappa_0=-\mu\alpha_0=\frac{\mu^2}{T}\sigma_0\longrightarrow \sigma_{\text{inc}}=\left(sT+\mu\rho\right)^2\sigma_0.
\end{equation}
As Lorentz symmetry relates the electric and heat currents to the momentum density by
\begin{equation}
\delta q^i=\delta \pi^i-\mu \delta j^i
\end{equation}
the incoherent current may be written
\begin{equation}
\delta j_\text{inc} = (sT+\mu\rho)\delta j-\rho\delta\pi
\end{equation}
which is proportional to the definition used in \cite{Davison:2015taa}. Moreover the momentum static susceptibility becomes
\begin{equation}
\pi^i=(\epsilon+p)\delta v^i+\ldots\quad \Rightarrow\quad\chi_{\pi\pi}=\epsilon+p
\end{equation}

Additional simplifications arise if conformal symmetry is not explicitly broken in the microscopic theory. In that case, we can write the pressure as
 \begin{equation}
 p=T^{d+1} f\left(\frac{T}{\mu}\right),
 \end{equation}
and therefore
 \begin{equation}
 \epsilon=d T^{d+1}f = d p\,,\quad \rho=-\frac{T^{d+2}}{\mu^2}f'\,,\quad s=(d+1)T^d f+\frac{T^{d+1}}{\mu}f'.
 \end{equation}
From this, we can derive the static susceptibilities
 \begin{equation}
 \label{eq:chisconf}
 \chi_{\epsilon\epsilon}=d(\epsilon+p)\,,\qquad \chi_{\epsilon\rho}=d\rho,
 \end{equation}
 which may be combined to obtain an expression for $\chi_{\text{inc}}$
 \begin{equation}
 \label{eq:chiincconf}
 \chi_{\inc}=\chi_{\rho\rho}-\frac{d\rho^2}{\epsilon+p}\,.
 \end{equation}
 Note that $\chi_{\text{inc}}(T\rightarrow0)\rightarrow0$, consistently with \eqref{LowTIncSusc}, as $\epsilon+p=\mu \rho$ and $\rho=\rho_0 \mu^d$ at $T=0$.

\subsection{Incoherent transport in holographic quantum critical metals \label{section:EMDIR}}

We now move on to analyse the transport properties of the translationally invariant systems governed by the IR QCPs of section \ref{section:QCPTI}. We start by calculating the optical conductivity at low frequencies $\omega\ll T$ and from this we extract the timescale $\tau_{eq}$ beyond which the system is governed by the hydrodynamic theory we have just presented. For the cases with a $z=1$ IR QCP, this timescale is parametrically longer than the inverse temperature $\tau_{eq}\gg T^{-1}$ and is directly controlled by the dangerously irrelevant deformation of the critical point that we emphasised in section \ref{sec:irrdefsubsec}.

At suitably late times, the hydrodynamic theory will still apply and we go on to calculate the values of certain thermodynamic and transport parameters in this theory when the system is near the IR QCP. Specifically, we evaluate the dc conductivity, static susceptibility and diffusivity of the `incoherent' charge which does not source momentum flow, and show that the diffusivity can be naturally expressed in units of the butterfly velocity and $\tau_{eq}$. We emphasize the difference in temperature scaling of these quantities depending on whether the QCP has $z=1$ or $z\ne1$, due to the importance of irrelevant deformations in the former case.

Finally, we discuss the low frequency scaling of the real part of the optical conductivity at the $T=0$ critical point. Our calculation improves previous results by carefully working out the dependence of this quantity on the irrelevant deformation for $z=1$ QCPs. This will allow us to resolve previous difficulties in understanding the general scaling properties of these QCPs in section \ref{section:MatchScaling}.  

\subsubsection{Optical conductivity at non-zero temperature and emergent long-lived excitation \label{section:accondm=0}}

The optical conductivity is defined from the usual holographic dictionary as
\begin{equation}
\label{eq:sigmadefn}
\sigma(\omega)\equiv-\frac{i}{\omega}\lim_{r\rightarrow0}\left(r^{2-d}\frac{a_x'(r)}{a_x(r)}\right),
\end{equation}
where $a_x(r,t)=a_x(r)e^{-i\omega t}$ is the perturbation of the gauge field along the $x$ direction that is ingoing at the horizon.

The near-horizon expansion of our solution is
\begin{equation}
\label{nearhorizonexpansionbackground}
\begin{split}
&D(r\to r_h)=4\pi T(r_h-r)+\ldots\,,B(r\to r_h)=1/(4\pi T(r_h-r))+\ldots\,,\\
&C(r\to r_h)=C_h+\ldots\,,\phi(r\to r_h)=\phi_h+\ldots\,,A(r\to r_h)=A_h(r_h-r)+\ldots
\end{split}
\end{equation}
The charge and entropy densities are given by the $r$-independent expressions:
\begin{equation}
\rho=-\frac{Z C^{d/2}A'}{\sqrt{BD}}=Z_h A_h C_h^{d/2}
\end{equation} and 
\begin{equation}
s=-\frac1T\left(\rho A-\frac{C^{1+d/2}(D/C)'}{\sqrt{BD}}\right)=4\pi C_h^{d/2}, 
\end{equation} 
where $Z_h\equiv Z(\phi(r_h))$.
We are mainly interested in the low $T$ solutions that reduce to \eqref{HVsol} in the IR as $T\rightarrow0$.

It is more convenient to use the rescaled perturbation
\begin{equation}
\tilde{a}_x\equiv \frac{a_x}{sT+\rho A},
\end{equation}
which obeys the equation of motion
\begin{equation}
\label{eq:tildeaxeqn}
\frac{d}{dr}\left[\sqrt{\frac{D}{B}}ZC^{d/2-1}\left(sT+\rho A\right)^2\tilde{a}_x'\right]+\omega^2\sqrt{\frac{B}{D}}ZC^{d/2-1}\left(sT+\rho A\right)^2\tilde{a}_x=0,
\end{equation}
in translationally invariant systems. For notational simplicity, we define
\begin{equation}
F(r)\equiv\sqrt{\frac{D}{B}},
\end{equation}
which has a zero at the finite temperature horizon $r=r_h$. 

To determine the low frequency behaviour of the conductivity, we follow the approach of \cite{Policastro:2002se,Herzog:2002fn} to solve the equation of motion \eqref{eq:tildeaxeqn} at small $\omega$. Near the horizon, the equation of motion becomes
\begin{equation}
\frac{d}{dr}\left[F(r)\tilde{a}_x'\right]+\frac{\omega^2}{F(r)}\tilde{a}_x=0,
\end{equation}
where $F(r\rightarrow r_h)\rightarrow 4\pi T(r_h-r)+\ldots$. The ingoing solution to this equation is
\begin{equation}
\tilde{a}_x=C_0\left(\frac{r_h-r}{r_h}\right)^{-i\frac{\omega}{4\pi T}}+\ldots.
\end{equation}
and we therefore make an ansatz
\begin{equation}
\tilde{a}_x\equiv \left(\frac{r_h-r}{r_h}\right)^{-i\frac{\omega}{4\pi T}}\mathcal{A}(r),
\end{equation}
for the gauge field perturbation, where $\mathcal{A}(r)$ is regular as $r\rightarrow r_h$. We then want to solve for $\mathcal{A}(r)$, which can be done perturbatively in $\omega$. That is, we expand
\begin{equation}
\mathcal{A}(r)=\mathcal{A}_0(r)+\left(\frac{\omega}{4\pi T}\right)\mathcal{A}_1(r)+\left(\frac{\omega}{4\pi T}\right)^2\mathcal{A}_2(r)+\ldots,
\end{equation}
so that at small frequencies
\begin{equation}
\tilde{a}_x(r)=\mathcal{A}_0(r)+\left(\frac{\omega}{4\pi T}\right)\left[\mathcal{A}_1(r)-i\mathcal{A}_0(r)\log\left(\frac{r_h-r}{r_h}\right)\right]+O(\omega^2).
\end{equation}

Near the horizon, where we must be sure that our solution is ingoing, the $\log$ term becomes very large. We can therefore only trust our solution for suitably small $\omega$. Quantitatively, for an imaginary $\omega=-i\tau_{eq}^{-1}$ this condition is $\tau_{eq}^{-1}\ll T$.

Bearing this in mind, we then solve the equation of motion order-by-order in $\omega$. At leading order,
\begin{equation}
\frac{d}{dr}\left[FZC^{d/2-1}\left(sT+\rho A\right)^2\mathcal{A}_0'\right]=0.
\end{equation}
The solution of this equation which is regular at the horizon is a constant: $\mathcal{A}_0(r)=c_0$.

At order $\omega$, the equation of motion is
\begin{equation}
\frac{d}{dr}\left[FZC^{d/2-1}\left(sT+\rho A\right)^2\left(\mathcal{A}_1-ic_0\log\left(\frac{r_h-r}{r_h}\right)\right)\right]=0,
\end{equation}
which has the general solution
\begin{equation}
\mathcal{A}_1(r)=c_2+ic_0\log\left(\frac{r_h-r}{r_h}\right)+c_3\int^r dr\frac{1}{FZC^{d/2-1}\left(sT+\rho A\right)^2}.
\end{equation}
To impose regularity at the horizon, it is first convenient to rewrite the log term as an integral
\begin{equation}
\mathcal{A}_1(r)=c_2-ic_0\log(r_h)+\int^r dr\left[\frac{c_3}{FZC^{d/2-1}\left(sT+\rho A\right)^2}-\frac{ic_0}{\left(r_h-r\right)}\right].
\end{equation}
Imposing regularity at the horizon gives
\begin{equation}
\mathcal{A}_1(r)=c_2-ic_0\log(r_h)-ic_0\int^r dr\left[-\frac{(4\pi T)Z_hC_h^{d/2-1}s^2T^2}{FZC^{d/2-1}\left(sT+\rho A\right)^2}+\frac{1}{\left(r_h-r\right)}\right],
\end{equation}
where subscript $h$ means evaluated on the horizon. Finally, we impose the boundary condition that $\mathcal{A}_i(r_h)=0$ for $i=1,2,\ldots$ to obtain
\begin{equation}
\mathcal{A}_1(r)=-ic_0\int^r_{r_h} dr\left[-\frac{(4\pi T)Z_hC_h^{d/2-1}s^2T^2}{FZC^{d/2-1}\left(sT+\rho A\right)^2}+\frac{1}{\left(r_h-r\right)}\right].
\end{equation}

Combining these results gives an expression for $\tilde{a}_x$ valid to order $\omega$. Expanding this solution near the boundary, we find that
\begin{equation}
\tilde{a}_x(r\rightarrow0)=c_0 \left(1-i\omega\tau_{eq}\right),
\end{equation}
where
\begin{equation}
\label{eq:taueqTI1}
\tau_{eq}=\int^{r_h}_{0} dr\left(\frac{Z_hC_h^{d/2-1}s^2T^2}{FZC^{d/2-1}\left(sT+\rho A\right)^2}-\frac{1}{4\pi T(r_h-r)}\right).
\end{equation}
Finally, substituting our solution into \eqref{eq:sigmadefn} yields the optical conductivity of the dual field theory
\begin{equation}
\label{eq:accondTI}
\sigma(\omega)=\frac{\rho^2}{(sT+\mu\rho)}\frac{i}\omega+\frac{s^2 T^2 Z_h}{(sT+\mu\rho)^2}\left(\frac{s}{4\pi}\right)^{\frac{d-2}d}\frac{1}{1-i\omega\tau_{eq}}\,.
\end{equation}
It is the sum of two terms. The first is a pole at $\omega=0$ due to momentum conservation, and the second is the contribution of processes that do not drag momentum. 

In the hydrodynamic theory of section \ref{section:thermohydro} the latter term is $\omega$-independent (equation \eqref{hydrocond}), whereas here it is dominated by a purely relaxational collective mode with lifetime $\tau_{eq}$. It is therefore manifest that hydrodynamics is a good description of these systems only over timescales $t\gtrsim\tau_{eq}$, beyond which the effects of non-hydrodynamic modes are important. While we generically expect the breakdown of hydrodynamics at short enough times, at this stage the result \eqref{eq:accondTI} is purely formal, as we cannot trust it unless $\tau_{eq}$ is parametrically longer than the thermal timescale $T^{-1}$.

The final stage of our calculation is to show that near the $z=1$ QCPs \eqref{IrrDefz=1}, $\tau_{eq}$ is in fact parametrically long and therefore can reliably be computed. In these cases, we can quantitatively trust the result \eqref{eq:accondTI}, and the long lifetime of the non-hydrodynamic collective mode then gives rise to a sharp Drude-like peak in the real part of the optical conductivity. 

To show this, it is useful to use the gravitational equations to rewrite the equation \eqref{eq:taueqTI1} for the lifetime as
\begin{equation}
\label{eq:taueq1}
\tau_{eq}=-\frac{1}{4\pi T}\int^0_{r_h} d\tilde{r}\left[\alpha\frac{C(\tilde{r})}{D(\tilde{r})}\frac{d}{d\tilde{r}}\left(\frac{1}{sT+\rho A(\tilde{r})}\right)-\frac{1}{r_h-\tilde{r}}\right],
\end{equation}
where $\alpha=s^3T^3Z_h\rho^{-2}(s/4\pi)^{-2/d}$. While the integral is sensitive to the form of the spacetime at all values of $r$, the important point is that for solutions which flow to $z=1$ in the IR, the $T\rightarrow0$ limit of the integral is dominated by a singular contribution from the IR part of the spacetime. 

Explicitly, suppose we have a spacetime that approaches \eqref{IrrDefz=1} and \eqref{CouplingsIRzeq1} in the deep interior when $T\rightarrow0$. Turning on a very small temperature will result in an event horizon at $r=r_h\gg L$ (as in \eqref{HVsol}), but the spacetime over the range $L\ll r\ll r_h$ will still be given by \eqref{IrrDefz=1} and \eqref{CouplingsIRzeq1} to leading order at small $T$. To capture the contribution of this part of the spacetime to the integral, we will integrate over a region $r_{UV}<r<r_{IR}$ where $L\ll r_{UV}$ and $r_{IR} \ll r_h$. In this region, $C(r)/D(r)$ is a constant and so it is trivial to determine the contribution to the integral to be
\begin{equation}
\begin{aligned}
\tau_{eq}(T\rightarrow0)&=-\frac{1}{4\pi T}\left[\alpha\frac{L_x^2}{L_t^2}\left(\frac{1}{sT+\rho A(\tilde{r})}\right)+\log\left(r_h-\tilde{r}\right)\right]^{r_{UV}}_{r_{IR}}\\
&=-\frac{1}{4\pi T}\left(\rho\alpha\frac{L_x^2}{L_t^2}\frac{A(r_{IR})-A(r_{UV})}{\left(sT+\rho A(r_{IR})\right)\left(sT+\rho A(r_{UV})\right)}+\log\left(\frac{r_h-r_{UV}}{r_h-r_{IR}}\right)\right).
\end{aligned}
\end{equation}
To proceed further, we define the cutoffs to be $r_{IR}=r_h \varepsilon_{IR}$ and $r_{UV}=L \varepsilon_{UV}^{-1}$ where $\varepsilon_{IR}\ll1$, $\varepsilon_{UV}\ll1$ and $\varepsilon_{IR}\varepsilon_{UV}\gg (L/r_h)\sim T$. We can always go to sufficiently small $T$ such that the cutoffs will satisfy \footnote{\baselineskip0.5pt We are suppressing factors of $L_x,L_t,\tilde{L},Z_0$ which are finite in the $T\rightarrow0$ limit.} 
\begin{equation}
A_{0}^{2}T^{-2\Delta_{A_0}}\ll\varepsilon_{IR}^{-(\theta-d-1-2\Delta_{A_0})}\ll1,\quad\quad\quad\quad A_{0}^{-2}T^{d+1-\theta}\ll\varepsilon_{UV}^{-(\theta-d-1-2\Delta_{A_0})}\ll1,
\end{equation}
since $\Delta_{A_0}<0$ and $d-\theta+1>0$. This ensures that
\begin{equation}
\rho A(r_{IR})\ll sT\ll \rho A(r_{UV}),
\end{equation}
and hence the contribution from this part of the spacetime as $T\rightarrow0$ is
\begin{equation}
\tau_{eq}(T\rightarrow0)=\frac{1}{4\pi T}\left(\alpha\frac{L_x^2}{L_t^2}\frac{1}{sT}-\log\left(1+\varepsilon_{IR}+\ldots\right)\right).
\end{equation}
The first term is the important one: it is independent of the specific choice of cutoffs, and gives a singular contribution to the integral in the $T\rightarrow0$ limit. Ignoring the subleading logarithmic term which we cannot trust yields
\begin{equation}
\tau_{eq}(T\rightarrow0)=\frac{L_x^2}{L_t^2}\frac{\alpha}{4\pi sT^2}=\frac{s^2TZ_hL_x^2}{4\pi \rho^2 L_t^2}\left(\frac{s}{4\pi}\right)^{-2/d}.
\end{equation}
Evaluating $\tau_{eq}$ for the $z=1$ solutions, this becomes
\begin{equation}
\label{eq:taueqz=1TI}
\tau_{eq}=\frac{\tilde{L}(d+1-\theta)}{L_tZ_0(\theta-d-1-2\Delta_{A_0})^2}A_0^{-2}\left(\frac{r_h}{L}\right)^{1-2\Delta_{A_0}},
\end{equation}
which (recalling that $r_h\sim T^{-1}$) can schematically be rewritten
\begin{equation}
\label{eq:taueqz=1TIb}
\tau_{eq}\sim\frac1T\left(\frac{T^{\Delta_g}}{g}\right)^2\,,\quad g\sim A_0\,.
\end{equation}
Since $\Delta_g<0$ for an irrelevant deformation, we indeed always find that there is a collective mode with a parametrically large lifetime $\tau_{eq}\gg T^{-1}$ near the $z=1$ QCPs. The $T$ dependence of the lifetime is determined by the dimension of the dangerously irrelevant deformation of the critical point sourced by $g$.

On the other hand, for the $z\neq1$ QCPs, we expect $\tau_{eq}\sim T^{-1}$. This is apparent by sending $\Delta_g\rightarrow0$ in \eqref{eq:taueqz=1TIb} (since the irrelevant deformation we have just discussed becomes marginal in this limit). Evaluating \eqref{eq:taueqTI1} on the IR $z\neq1$ solutions, we indeed find a contribution $\tau_{eq}\sim 1/T$ consistent with this expectation.

%%%%%%%%%%%%%%%%%%%%%%%%%%%%%%%%%%%%%%%%%%%
\subsubsection{Incoherent conductivity, susceptibility and diffusivity near the QCP}

At suitably late times, systems governed by both types of IR QCPs ($z=1$ and $z\neq1$) are governed by the hydrodynamics of section \ref{section:thermohydro}. We will now show that the parameters governing the hydrodynamic transport of the incoherent charge ($\sigma_{\text{inc}},\chi_{\text{inc}}$ and $D_{\text{inc}}$) are determined by the particular IR QCP. The $z=1$ and $z\ne1$ cases are qualitatively different in that the irrelevant deformation sourced by $A_0$ plays a vital role in determining the parameters near the $z=1$ QCPs.\\\strut\\

\paragraph{dc incoherent conductivity\\\strut\\}

For static, radially dependent solutions of the theory \eqref{actionEMD}, the incoherent dc conductivity $\sigma_{\text{inc}}$ can be expressed exactly in terms of the gravitational solution at the horizon \cite{Hartnoll:2007ip,Jain:2010ip,Chakrabarti:2010xy,Hartnoll:2015sea}
\begin{equation}
\label{inccondhorizon}
\sigma_{\text{inc}}=(sT+\mu\rho)^2\sigma_o=\left(s T\right)^2 \left(\frac{s}{4\pi}\right)^{(d-2)/d} Z(\phi_h).
\end{equation}

As a consequence of this horizon formula, for the theories we described in section \ref{section:QCPTI} $\sigma_{\text{inc}}$ is directly sensitive to the IR solutions that capture the quantum critical physics. Evaluating \eqref{inccondhorizon} for these IR solutions, we find that
\begin{equation}
\label{eq:sigmascaleholo1}
\sigma_{\text{inc}}=\left\{\begin{array}{lc}
{L_t^2 L_x^{3d-2}}{\tilde L^{-2}}Z_0(d+z-\theta)^2\left({r_h}/{L}\right)^{2-2z-d+\theta}\,,&z\neq1\,,
\\{L_t^2 L_x^{3d-2}}{\tilde L^{-2}}Z_0 (d+1-\theta)^2\left({r_h}/{L}\right)^{\theta-d-2\Delta_{A_0}}\,,&z=1\,.
\end{array}\right.
\end{equation}
This leads to a low temperature scaling
\begin{equation}
\label{eq:sigmascaleholo}
\sigma_{\text{inc}}\sim T^{2+\frac{3d-3\theta-2+2\Phi}{z}}\sim T^{2+\frac{d-2-\theta+2\Delta_{A_0}}{z}},
\end{equation}
where we are suppressing numerical prefactors and the dependence on the dimensionless IR parameters $L_x,L_t$ etc. Importantly, there is no dependence on the value of the irrelevant coupling $A_0$. \\\strut\\

\paragraph{Incoherent susceptibility\label{subsec:incsus}\\\strut\\}

At low temperatures, the incoherent susceptibility of the states we are studying can also be explicitly written in terms of the near-horizon gravitational solution that captures the quantum critical physics. This is manifest from the equation \eqref{LowTIncSusc} derived earlier, that relates $\chi_{\text{inc}}$ to the heat capacity $c_\rho$ at low temperatures. At low temperatures, our solutions have heat capacity $c_\rho\sim T^{(d-\theta)/z}$ and thus
\begin{equation}
\label{eq:chiscaleholo}
\chi_{\text{inc}} = T\rho^2 c_\rho \sim A_0^2\,T^{\frac{d-\theta+z}{z}},
\end{equation}
in this limit. The factor of $A_0^2$ comes from the factor of $\rho^2$ in the relation between $\chi_{\text{inc}}$ and $c_\rho$. The result \eqref{eq:chiscaleholo} is true for both types of IR solution. For the $z=1$ solutions, $A_0$ is an irrelevant coupling, and $\chi_\text{inc}$ then directly depends on this coupling. More precisely, evaluating the low temperature incoherent susceptibility \eqref{LowTIncSusc} on the solutions of section \ref{section:QCPTI} gives
\begin{equation}
\label{eq:chiholo}
\chi_{\text{inc}}=\left\{\begin{array}{lc}
{L_t L_x^{3d}}{\tilde L^{-3}}Z_0\frac{z-1}{z}(d-\theta)(d+z-\theta)^2\left({r_h}/{L}\right)^{\theta-d-z}\,,&z\neq1\,,
\\{L_t L_x^{3d}}{\tilde L^{-3}}Z_0^2 A_0^2(\theta-d-1-2\Delta_{A_0})^2 (d-\theta)(d+1-\theta)\left({r_h}/{L}\right)^{\theta-d-1}\,,&z=1\,.
\end{array}\right.
\end{equation}

The temperature scaling for the incoherent susceptibility in equation \eqref{eq:chiscaleholo} is not necessarily true for $z=\infty$ states with a finite zero temperature entropy density (i.e.~for IR solutions with a metric that is AdS$_2\times$ $R^d$), as in these cases $c_\rho$ is sensitive to irrelevant deformations. The results in the following subsection also do not necessarily apply to these cases, which we will not discuss in generality as they were studied in detail in \cite{Blake:2016jnn}. 

\subsubsection{Diffusivity near the QCP \label{subsection:m=0Diffusivity}}

By combining our results for the incoherent dc conductivity and susceptibility, we can now determine the low temperature limit of the incoherent diffusivity $D$ using the Einstein relation \eqref{EinsteinInc}. As explained in section \ref{subsection:hydrowithE}, this is the same as the low temperature limit of the thermal diffusivity $D_T=\kappa/c_\rho$. As a consequence of the results in the preceding subsections, both diffusivities may explicitly be written in terms of the gravitational solution near the horizon.

Combining the results \eqref{eq:sigmascaleholo} and \eqref{eq:chiscaleholo}, the dependence of the diffusivity $D$ on $T$ and $A_0$ near the quantum critical point is given by
\begin{equation}
D=\frac{\sigma_{\text{inc}}}{\chi_{\text{inc}}}\sim A_0^{-2}T^{1+\frac{2\left(\Delta_{A_0}-1\right)}{z}}.
\end{equation}
For the IR solutions where $A_0$ is an irrelevant coupling (i.e.~the $z=1$ solutions), the low temperature diffusivity manifestly depends on the irrelevant coupling.

Following \cite{Blake:2016sud,Blake:2016wvh,Lucas:2016yfl,Baggioli:2016pia,Blake:2016jnn,Blake:2017qgd} it is instructive to express the incoherent diffusivity in units of $v_B^2\tau_L$, where $v_B$ is the quantum butterfly velocity (that characterizes the speed at which quantum chaos spreads) and the `Planckian' or Lyapunov timescale $\tau_L=\hbar/(2\pi k_BT)$. $v_B$ and $\tau_L$ are near-horizon observables which provide an IR velocity and timescale that scale near the critical point as \cite{Shenker:2013pqa,Maldacena:2015waa,Blake:2016wvh,Roberts:2016wdl}
\begin{equation}
v_B^2\sim T^{2-\frac{2}{z}},\;\;\;\;\;\;\;\;\;\;\;\;\;\;\;\; \tau_L\sim T^{-1}, 
\end{equation}
and are independent of $A_0$. Expressed in these units, the low temperature diffusivity scales as
\begin{equation}
D\sim A_0^{-2} T^\frac{2\Delta_{A_0}}{z} v_B^2\tau_L.
\end{equation}
This expression is valid for both types of IR solution \footnote{\baselineskip0.5pt It does not apply to the AdS$_2\times$R$^d$ cases, for the reasons described in the previous section. See \cite{Blake:2016jnn} for a discussion of diffusion in such solutions.}.

For the first type of IR solution ($z\neq1$), $D/v_B^2\tau_L$ is temperature-independent at low $T$, and can therefore be expressed solely in terms of the marginal parameters that characterise the IR fixed point ($L_t,A_0$ etc.). Computing the coefficient explicitly, we find that
\begin{equation}
\label{IncDifBoundSat}
D=\frac{z}{2(z-1)}v_B^2\tau_L.
\end{equation}
for these IR solutions. The coefficient is actually independent of all of the marginal parameters characterising the fixed point, and depends only on the dynamical critical exponent $z$. The coefficient is the same as for the low temperature thermal diffusivity in states where translational symmetry is broken \cite{Blake:2017qgd}. In fact, the result (\ref{IncDifBoundSat}) can be understood as a limiting case of the results of \cite{Blake:2017qgd}. The simplest way to see this is to note that, in the translationally invariant limit, the holographic expression for the open circuit dc thermal conductivity $\kappa$ (equation (2.5) of \cite{Blake:2017qgd}) is finite and in agreement with our equations (\ref{eq:kappahydro}) and (\ref{inccondhorizon}). As the heat capacity $c_\rho$ is continuous in this limit, the thermal diffusivity $D_T\equiv\kappa/c_\rho$ must also be. Finally, in the low temperature limit where we are working, the incoherent diffusivity $D$ is equal to the thermal diffusivity (equation (\ref{eq:relationbetweenDs})), and from this (\ref{IncDifBoundSat}) follows.

For the second type of IR solution ($z=1$), $\tau_L$ does not appear to be the appropriate timescale that sets the diffusivity $D$. In \cite{Hartman:2017hhp,Lucas:2017ibu} it was argued that the appropriate timescale is in fact $\tau_{eq}$. Since $\tau_{eq}\sim T^{-1}\sim\tau_L$ for $z\ne1$ solutions, the result \eqref{IncDifBoundSat} for the $z\ne1$ cases is consistent with this. But as we discussed above, $\tau_{eq}$ is parametrically larger than $T^{-1}$ for the $z=1$ cases, and using the explicit expression \eqref{eq:taueqz=1TI} we find that the diffusivity can be written at low temperatures as
\begin{equation}
\label{eq:Dirr}
D=\frac{2}{d+1-\theta}\;v_B^2 \tau_{eq}\,,
\end{equation}
for these cases. While both $D$ and $\tau_{eq}$ are larger than one would naively expect due to their dependence on the dangerously irrelevant coupling $A_0$, $D/v_B^2\tau_{eq}$ is independent of $T$ and $A_0$ and is given by a universal number that depends only on the exponents of the fixed point. 

\subsubsection{ac conductivity \label{section:accondnonzerodensity}}

Until now we have considered the quantum critical dynamics in the range $\omega\ll T$. For completeness, we now consider the opposite limit $\omega\gg T$ i.e.~exactly at the $T=0$ critical point. As we have established that the dynamics of $\delta\rho_{\text{inc}}$ are sensitive to the nature of the IR QCP, we will study its zero temperature ac conductivity $\sigma_{\text{inc}}(\omega)$. Defining it via a Kubo formula, it is related to the thermoelectric conductivities by
\begin{equation}
\sigma_{\text{inc}}(\omega)\equiv\lim_{k\rightarrow0}\frac{i}{\omega}G^R_{j_{\text{inc}}j_{\text{inc}}}(\omega,k)=s^2T^2\sigma(\omega)-2\rho sT^2\alpha(\omega)+\rho^2\bar{\kappa}(\omega).
\end{equation}
Implementing a Ward identity between the thermoelectric conductivities that is required by the UV Lorentz symmetry of holographic theories \cite{Herzog:2009xv} we then find that it is related to the electrical conductivity $\sigma$ by
\begin{equation}
\text{Re}\;\sigma_{\text{inc}}(\omega)=(sT+\rho\mu)^2\;\text{Re}\;\sigma(\omega).
\end{equation}
Taking the real part is necessary because the electrical conductivity $\sigma(\omega)$ has an extra imaginary piece which diverges as $\omega\rightarrow0$ due to translational symmetry. Taking the zero temperature limit, we find that
\begin{equation}
\label{eq:optsigincdef}
\text{Re}\;\sigma_{\text{inc}}(\omega,T=0)=\rho^2\mu^2\;\text{Re}\;\sigma(\omega,T=0).
\end{equation}
At low $\omega$, the $\omega$ dependence of the quantity on the right hand side depends only on the exponents characterising the IR solution \cite{Charmousis:2010zz,Gouteraux:2011ce,Gouteraux:2013oca}. By performing a more careful matching calculation and keeping track of the overall normalisation, we are going to show that in fact the quantity on the left hand side is a more natural quantity as we can write its low frequency limit (including the prefactor) exactly in terms of the parameters of the IR solution. 

To show this, we study the equation of motion for the spatially uniform perturbation perturbation of the U(1) gauge field $a_x$ (in the full spacetime), which is
\begin{equation}
\label{eq:gaugeeom}
\frac{d}{dr}\left[ZC^{\frac{d-2}{2}}\sqrt{\frac{D}{B}}a_x'\right]+ZC^{\frac{d-2}2}\left[\sqrt{\frac{B}D}\omega^2-\frac{A'^2Z}{\sqrt{BD}}\right]a_x=0\,.
\end{equation}
It is convenient to first change variables to
\begin{equation}
\bar{a}=\sqrt{\bar Z}a_x\,,\quad\quad\quad\bar Z=Z C^{\frac{d-2}{2}},\,\quad\quad\quad \frac{d\bar r}{dr}=\sqrt{\frac{B}D},
\end{equation}
so that this equation can be written in the Schrödinger form
\begin{equation}
\label{Seq}
\frac{d^2\bar{a}}{d\bar r^2}+(\omega^2-V_s)\bar{a}=0,\quad\quad\quad V_s=\frac{A'^2Z}{B}+\frac{(\partial_{\bar r}\bar Z)^2}{4\bar Z^2}+\frac12\left(\partial_{\bar r}\right)^2\ln\bar Z.
\end{equation}
In this form, it is easy to prove that there is a radially conserved quantity
\begin{equation}
\label{consflux}
\mathcal F\equiv i\left(\bar{a}^\star\partial_{\bar r}\bar{a}-\bar{a}\partial_{\bar r}\bar{a}^{\star}\right),
\end{equation}
for real values of $\omega$. Evaluating it near the boundary, we find that $\mathcal{F}$ is related to the electrical conductivity (up to a constant prefactor) by
\begin{equation}
\label{eq:accondflux}
\text{Re}\;\sigma(\omega)=\frac{\mathcal{F}}{\omega \left|a_x^{(0)}\right|^2},
\end{equation}
where $a_x^{(0)}$ is the value of $a_x$ at the AdS boundary, which we can set to $1$ by the scaling symmetry of the perturbation equation. 

To determine the conductivity, we will now use a matching argument \cite{Gubser:2008wz} to calculate $\mathcal F$ in the IR region of the geometry for a solution $a_x$ which is ingoing at the horizon and whose asymptotic value is $1$. In the zero temperature IR solution that is valid in the deep interior, the Schrödinger potential is $V_s=V_{s,0}/(4\bar r^2)$, where the value of $V_{s,0}$ depends on the class of IR solution:
\begin{equation}
\label{Vo}
\begin{split}
z\neq1\,:&\quad V_{s,0}=\frac{(d-\theta-2+2z)(d-2-\theta+4z)}{z^2},\\
z=1\,:&\quad  V_{s,0}=(d-\theta-2\Delta_{a_0})(d+2-\theta-2\Delta_{a_0}).
\end{split}
\end{equation}
For the $z=1$ solutions we neglected the $\sim {A'}^2$ contribution to $V_s$, which is subleading in the IR. Neglecting these will allow us to do a consistent matching of solutions. The ingoing solution of the Schrödinger equation \eqref{Seq} with this potential is
\begin{equation}
\label{eq:Hankelsol}
\bar{a}=\bar{a}_0 \sqrt{\bar r} H_\nu^{(1)}\left(\omega\bar r\right),\quad 2\nu=\sqrt{1+ V_{s,0}}=\left\{\begin{array}{ll}\frac{d-2+3z-\theta}{z}\;\;&z\neq1\\d+1-\theta-2\Delta_{A_0}\;\;&z=1 \end{array}\right.\,,
\end{equation}
where $\bar{a}_0$ is a constant and $H_\nu^{(1)}$ is the Hankel function of the first kind. Near the boundary of the IR region of the spacetime, this solution has the form
\begin{equation}
\label{eq:IRsolnexp}
a_x(r)= a_0 r^{\frac{z+\zeta-2}2}\left(r^{-z\nu}+\#\;\omega^{2\nu}r^{+z\nu}+\ldots\right)\sim\left\{\begin{array}{ll} a_0\left(r^{-(d+z-\theta)}+\#\;\omega^{3+\frac{d-\theta-2}{z}}r^{2(z-1)}+\ldots\right)&z\neq1\\a_0\left(r^{2\Delta_{A_0}+\theta-d-1}+\#\;\omega^{d+1-\theta-2\Delta_{A_0}}r^0+\ldots\right)\;\;&z=1 \end{array}\right.
\end{equation}
where $a_0\sim \omega^\nu \bar{a}_0$ is a constant and $\#$ denote complex constants that depend on the parameters of the IR solution but not on the irrelevant coupling $A_0$. Evaluating the radially conserved quantity $\mathcal F$ yields the conductivity
\begin{equation}
\text{Re}\;\sigma(\omega,T=0)\sim\left|a_0\right|^2 \omega^{2\nu-1}\sim\left\{\begin{array}{ll}\left|a_0\right|^2 \omega^{2+\frac{d-\theta-2}{z}},&z\neq1\\\left|a_0\right|^2 \omega^{d-\theta-2\Delta_{A_0}}\;\;&z=1 \end{array}\right.
\end{equation}

The final step of the calculation is to fix the overall normalisation constant $a_0$ such that this solution asymptotes to $1$ at the boundary of the AdS spacetime. One might expect this normalisation constant to depend on the full RG flow of the theory, and it does, but exactly in such a way that $\sigma_{\text{inc}}(\omega)$ can be written simply in terms of the parameters of the IR solution.

The expansion \eqref{eq:IRsolnexp} is valid when $1\ll\bar{r}\ll\omega^{-1}$ (but the solution \eqref{eq:Hankelsol} is valid in the wider region $1\ll\bar{r}<\infty$). To fix the constant, we will match it to a solution which is valid in this region, but is also valid near the AdS boundary. To find this second solution, we set $\omega=0$ in the equation of motion \eqref{eq:gaugeeom} and solve to obtain
\begin{equation}
\label{eq:com}
a_x(r)=\frac{A(r)}{\mu}+a_1A(r)\int ^r_{\text{bdy}}dr\left(\sqrt{\frac{B}{D}}\frac{C^{1-d/2}}{ZA^2}\right),
\end{equation}
where $a_1$ is a constant and we have normalised the solution so that the boundary value $a_x^{(0)}=1$. This solution is valid for $0< r\ll\omega^{-1}$. We can expand this in the deep interior (where the integrand diverges) to obtain the following solution which is valid in the region $1\ll\bar{r}\ll\omega^{-1}$
\begin{equation}
\label{eq:matching2}
\begin{split}
z\neq1\,:&\quad a_x(r)\sim\frac{A_0}{\mu}r^{-(d+z-\theta)} + a_1r^{2(z-1)} ,\\
z=1\,:&\quad a_x(r)\sim\frac{A_0}{\mu}r^{\zeta-1}+a_1 r^0,
\end{split}
\end{equation}
where we have suppressed the dependence on the parameters of the IR solution except the irrelevant coupling $A_0$. For each power of $r$, we keep only the leading term in a small $a_1$ expansion, as higher order terms turn out to be suppressed at low frequencies. As both \eqref{eq:IRsolnexp} and \eqref{eq:matching2} are valid over a region $1\ll\bar{r}\ll\omega^{-1}$ of the radial co-ordinate which is parametrically large at small $\omega$, we can match them in this region to fix the constants $a_0$ and $a_1$. This yields $a_0\sim A_0/\mu$ and thus the small $\omega$ conductivity is
\begin{equation}
\begin{split}
z\neq1\,:&\quad \sigma(\omega,T=0)\sim \frac{A_0^2}{\mu^2}\omega^{2+\frac{d-\theta-2}{z}},\\
z=1\,:&\quad \sigma(\omega,T=0)\sim\frac{A_0^2}{\mu^2}\omega^{d-\theta-2\Delta_{A_0}}.
\end{split}
\end{equation}
This object is awkward because it depends on the chemical potential $\mu$, which cannot be expressed in a simple way in terms of the IR solution. However, using the result \eqref{eq:optsigincdef} for $\sigma_{\text{inc}}(\omega,T=0)$, we see that the powers of $\mu$ cancel and that, as in the dc limit, $\sigma_{\text{inc}}$ can be written simply in terms of the the parameters in the IR solution. Its dependence on $\omega$ and the irrelevant coupling $A_0$ can therefore easily be extracted and written as
\begin{equation}
\label{eq:sigmaincholorest}
\begin{split}
z\neq1\,:&\quad \sigma_{\text{inc}}(\omega,T=0)\sim \omega^{2+\frac{d-2-\theta}{z}},\\
z=1\,:&\quad \sigma_{\text{inc}}(\omega,T=0)\sim A_0^4\;\omega^{d-\theta-2\Delta_{A_0}}.
\end{split}
\end{equation}
Finally, we note that the numerical prefactors and dependence of $\sigma_{\text{inc}}(\omega,T=0)$ on the other IR parameters ($L_t,L_x$ etc.) can be calculated explicitly by the procedure we have just described. As we are only interested in the overall scaling behaviour, we have omitted these details for clarity.

Note that the scaling with $\omega$ in \eqref{eq:sigmaincholorest} does not match the temperature scaling of the incoherent conductivity in \eqref{eq:sigmascaleholo} when $z=1$ and $\Delta_{A_0}\neq0$, due to the factors of the irrelevant coupling $A_0$. We will return to the failure of naive scaling theory in section \ref{section:MatchScaling}.

\section{Zero density case \label{section:transportmomrel}}

We will now address the second class of holographic systems that we introduced in section \ref{section:QCPTI}: those at zero density but where translational symmetry is broken by the massless scalar fields $\psi_i$. Although these systems are in many ways different from those we looked at in the previous section, they share the property that their local equilibration time $\tau_{eq}$ can be parametrically longer than the inverse temperature due to its sensitivity to a dangerously irrelevant deformation of the quantum critical point. The physical origin of the long-lived excitation in these systems is clear: the irrelevant deformation breaks the translational symmetry of the QCP, leading to the relaxation of the system's momentum over a long timescale $\tau_{eq}$.

\subsection{dc transport in quantum critical phases breaking translations}

We firstly review the late time transport properties of the zero density systems without translational symmetry introduced in section \ref{section:QCPTI}. Their dc thermal conductivity is given by \cite{Donos:2014cya}:
\begin{equation}
\label{dceqholo}
\bar\kappa=\frac{4\pi s T}{m^2 Y_h}
\end{equation}
where $Y_h$ is the value of $Y(\phi)$ at the event horizon. This expression is nonperturbative in $m$ and so is valid independent of whether the translational symmetry breaking caused by $m$ results in the total momentum of the system relaxing quickly or slowly. Independently of how quickly momentum relaxes, over sufficiently long times the momentum will have relaxed and we expect perturbations in the energy and heat density to diffuse at the thermal diffusivity
\begin{equation}
D_T=\frac{\bar\kappa}{c_T},
\end{equation}
where $c_T=T ds/dT$ is the heat capacity (see e.g.~\cite{Davison:2014lua}).

For cases when the breaking of translational symmetry is caused by a deformation that is marginal near the QCP ($z\neq1$), the thermal diffusivity is related in a very simple way to the butterfly velocity and the thermal timescale \cite{Blake:2016jnn,Blake:2016sud,Blake:2017qgd}
\begin{equation}
D_T=\frac{z}{2(z-1)}\frac{v_B^2}{2\pi T}\,.
\end{equation}
This relation breaks down in the case $z=1$, as the thermal diffusivity becomes anomalously large due to its sensitivity to the irrelevant deformation that breaks translational symmetry \cite{Blake:2017qgd}. More precisely, when $z=1$ we find that
\begin{equation}
D_T=v_B^2 \frac{L_x^2(d+1-\theta)}{\tilde L L_t Y_0 m^2}\left(\frac{L}{r_h}\right)^{2\Delta_m-1},
\end{equation}
so that the thermal diffusivity in units of the butterfly velocity is controlled by the timescale
\begin{equation}
\tau\sim\frac1{m^2}\left(\frac{L}{r_h}\right)^{2\Delta_m-1}\sim\frac1T\left(\frac{T^{\Delta_m}}{m}\right)^2,
\end{equation}
which is parametrically longer than $T^{-1}$ since $\Delta_m<0$ always.

In the following subsection we are going to show that this timescale is in fact the lifetime of perturbations of the total momentum of the system. These excitations have a long lifetime as it is an irrelevant deformation of the IR theory that breaks translational symmetry \cite{Hartnoll:2012rj}. More precisely, we will show that
\begin{equation}
D_T=\frac{2}{\left(d+\theta-1\right)}v_B^2\tau_{eq},
\end{equation}
where $\tau_{eq}$ is the lifetime of the Drude-like excitation in the ac thermal conductivity
\begin{equation}
\tau_{eq}=\frac{L_x^2(d+1-\theta)}{\tilde L L_t Y_0 m^2}\left(\frac{L}{r_h}\right)^{2\Delta_m-1}.
\end{equation}

\subsection{ac thermal conductivity and long-lived excitation}

The calculation of the ac heat conductivity $\bar{\kappa}(\omega)$, and the lifetime $\tau_{eq}$ of its longest-lived pole, is mathematically very similar to the calculation of the optical conductivity that we presented in section \ref{section:accondm=0} and so we will be brief in some manipulations. We will make repeated use of the Einstein equation
\begin{equation}
\label{eq:axionbgrel}
\frac{d}{dr}\left[\frac{C^{d/2+1}}{\sqrt{BD}}\left(\frac{D}{C}\right)'\right]=m^2\sqrt{BD}C^{d/2-1}Y,
\end{equation}
in these calculations. Integrating \eqref{eq:axionbgrel} from the horizon to the boundary gives
\begin{equation}
\frac{C^{d/2+1}}{\sqrt{BD}}\left(\frac{D}{C}\right)'=-sT-m^2\int^{r_h}_{r}dr\sqrt{BD}C^{d/2-1}Y\,,
\end{equation}
where we used the near horizon expansion \eqref{nearhorizonexpansionbackground} (setting the gauge field to zero).
To determine the heat conductivity, we solve the coupled equations of motion for spatially uniform perturbations of the metric and the massless scalar field $\delta g^x_t$, $\delta g^x_r$ and $\delta\psi_x$ \footnote{\baselineskip0.5pt Indices are raised with the background metric.}. It is convenient to work with the variable
\begin{equation}
\Pi_x\equiv-\frac{{\delta g^x_t}'+i\omega \delta g^x_r}{\left(D/C\right)'},
\end{equation}
which obeys the equation of motion
\begin{equation}
\frac{d}{dr}\left[\sqrt{\frac{D}{B}}\frac{1}{YC^{d/2}}\left(\frac{C^{d/2+1}}{\sqrt{BD}}\left(\frac{D}{C}\right)'\right)^2\Pi_x'\right]+\omega^2\sqrt{\frac{B}{D}}\frac{1}{YC^{d/2}}\left(\frac{C^{d/2+1}}{\sqrt{BD}}\left(\frac{D}{C}\right)'\right)^2\Pi_x=0.
\end{equation}
Utilising the background relation \eqref{eq:axionbgrel}, this equation can be written in a more useful form as 
\begin{equation}
\begin{aligned}
\label{eq:axionfluceq}
&\frac{d}{dr}\left[\sqrt{\frac{D}{B}}\frac{1}{YC^{d/2}}\left(sT+m^2\int^{r_h}_rdr\sqrt{BD}C^{d/2-1}Y\right)^2{\Pi}_x'\right]\\
&+\omega^2\sqrt{\frac{B}{D}}\frac{1}{YC^{d/2}}\left(sT+m^2\int^{r_h}_rdr\sqrt{BD}C^{d/2-1}Y\right)^2{\Pi}_x=0.
\end{aligned}
\end{equation}
The heat conductivity $\bar{\kappa}(\omega)$ is given by
\begin{equation}
\label{eq:kappadefholo}
\bar\kappa(\omega)=-\frac{i m^2}{\omega T}\frac{1}{(d-1)}\frac{\Pi_x^{(d-1)}}{\Pi_x^{(0)}},
\end{equation}
where $\Pi_x$ is ingoing at the horizon and $\Pi^{(i)}_x$ denotes the coefficient of $r^i$ in the near-boundary expansion of $\Pi_x$. 

As in section \ref{section:accondm=0}, we make an ingoing ansatz for $\Pi_x$, and then solve the equation of motion \eqref{eq:axionfluceq} perturbatively at small $\omega$. Up to $O(\omega)$, the equation \eqref{eq:axionfluceq} is a total derivative and so can trivially be integrated to give (for constant $c_0$)
\begin{equation}
\Pi_x=\left(\frac{r_h-r}{r_h}\right)^{-i\frac{\omega}{4\pi T}}\left[1+\frac{i\omega}{4\pi T}\int ^r_{r_h}dr\left(\frac{\frac{4\pi s^2T^3}{Y_hC_h^{d/2}}YC^{d/2}\sqrt{\frac{D}{B}}}{\left(sT+m^2\int^{r_h}_rdr\sqrt{BD}C^{d/2-1}Y\right)^2}-\frac{1}{r_h-r}\right)\right].
\end{equation}
As before, we can trust this solution for frequencies $\omega\ll T$.

From this we can extract the $\bar{\kappa}(\omega)$ using equation \eqref{eq:kappadefholo} to find
\begin{equation}
\label{eq:holoheatcondresults}
\bar\kappa(\omega)=\frac{\bar\kappa_{dc}}{1-i\omega\tau_{eq}}\,,\qquad \bar\kappa_{dc}=\frac{4\pi s T}{m^2 Y_h},
\end{equation}
where $\tau_{eq}$ is given by the integral
\begin{equation}
\begin{aligned}
\label{eq:taum!=0}
\tau_{eq}&=\int^{r_h}_0dr\left[\frac{s^2T^2YC^{d/2}}{Y_hC_h^{d/2}\sqrt{\frac{D}{B}}\left(sT+m^2\int^{r_h}_{r}dr\sqrt{BD}C^{d/2-1}Y\right)^2}-\frac{1}{4\pi T (r_h-r)}\right]\\
&=\int^{r_h}_0dr\left(\frac{4\pi sT^2YC^{d/2}}{m^2Y_h}\frac{C}{D}\frac{d}{dr}\left[\frac{1}{\left(sT+m^2\int^{r_h}_{r}dr\sqrt{BD}C^{d/2-1}Y\right)}\right]-\frac{1}{4\pi T (r_h-r)}\right).
\end{aligned}
\end{equation}
The dc value of $\bar{\kappa}(\omega)$ matches \eqref{dceqholo} as expected.

$\tau_{eq}$ formally gives the lifetime of the longest-lived non-hydrodynamic excitation of the theory. But we can only trust our results \eqref{eq:holoheatcondresults} and \eqref{eq:taum!=0} for parametrically large $\tau_{eq}\gg T^{-1}$. Evaluated on the solutions described in \cite{Andrade:2013gsa,Davison:2014lua} returns $\tau_{eq}=4\pi T/m^2$, which matches the result in \cite{Davison:2014lua}.

We will now evaluate $\tau_{eq}$ for solutions that flow in the IR to the $z=1$ geometries \eqref{IrrDefz=1} and show that in these cases $\tau_{eq}$ is parametrically long due to its sensitivity to the irrelevant deformation that breaks translational symmetry. As in section \ref{section:accondm=0}, the important point is again that, in these cases, the integral is dominated by the contribution of a region of the spacetime close to the horizon. Since the mathematical manipulations are very similar to those of section \ref{section:accondm=0}, we will be brief.

We consider the contribution to the integral of the region $r_{UV}<r<r_{IR}$ where $L\ll r_{UV}$ and $r_{IR}\ll r_h$. At leading order at small $T$, the solution in this region will be given by \eqref{IrrDefz=1} and \eqref{CouplingsIRzeq1m} and thus $C(r)/D(r)=L_x^2/L_t^2$ is a constant. It is then trivial to perform the integral over this region to obtain
\begin{equation}
\begin{aligned}
\tau_{eq}(T\rightarrow0)=&\frac{4\pi sT^2L_x^2}{Y_hL_t^2}\frac{\int^{r_{IR}}_{r_{UV}}dr\sqrt{BD}C^{d/2-1}Y}{\left(sT+m^2\int^{r_h}_{r_{IR}}dr\sqrt{BD}C^{d/2-1}Y\right)\left(sT+m^2\int^{r_h}_{r_{UV}}dr\sqrt{BD}C^{d/2-1}Y\right)}\\
&+\frac{1}{4\pi T}\log\left(\frac{r_h-r_{IR}}{r_h-r_{UV}}\right).
\end{aligned}
\end{equation}
We define the cutoffs to be $r_{IR}=\varepsilon_{IR}r_h$ and $r_{UV}=\varepsilon_{UV}^{-1}L$ with $\varepsilon\ll1$ and $\varepsilon_{IR}\varepsilon_{UV}\gg L/r_h\sim T$. We can always go to a sufficiently low $T$ such that the cutoffs satisfy the identities
\begin{equation}
m^2 T^{-2\Delta_m}\ll\varepsilon_{IR}^{d+1-\theta-2\Delta_m}\ll1,\quad\quad\quad\quad m^{-2}T^{d+1-\theta}\ll\varepsilon_{UV}^{d+1-\theta-2\Delta_m}\ll 1,
\end{equation}
since $\Delta_m<0$ and $d+1-\theta>0$. Within this part of the spacetime,
\begin{equation}
m^2\int^{r_h}_{r_{IR}}dr\sqrt{BD}C^{d/2-1}Y\ll sT\ll m^2\int^{r_{IR}}_{r_{UV}}dr\sqrt{BD}YC^{d/2-1},
\end{equation}
and thus the contribution to $\tau_{eq}$ from this region is 
\begin{equation}
\tau_{eq}(T\rightarrow0)=\frac{4\pi TL_x^2}{Y_hL_t^2m^2}+\frac{1}{4\pi T}\log(1-\varepsilon_{IR}).
\end{equation}
The $\log$ term is subdominant as $T\rightarrow0$ and so the result in this limit is
\begin{equation}
\label{TauEqHoloM}
\tau_{eq}=\frac{L_x^2(d+1-\theta)}{\tilde L L_t Y_0 m^2}\left(\frac{L}{r_h}\right)^{2\Delta_m-1}\sim \frac1T\left(\frac{T^{\Delta_m}}{m}\right)^2.
\end{equation}
This satisfies $\tau_{eq}\gg T^{-1}$ as advertised, since $\Delta_m<0$ for an irrelevant deformation.

\subsection{Zero temperature conductivity \label{section:accondzerodensity}}

In the previous section, we computed the low (zero) frequency dependence of the thermal conductivity at nonzero temperature. We would now like to take the opposite order of limits $T\ll\omega$ and compute how the thermal conductivity depends on the irrelevant deformation.

To this end, we repeat the calculation of section \ref{section:accondnonzerodensity}. The relevant equation to solve is again equation \eqref{eq:axionfluceq} from which we can extract the heat conductivity via \eqref{eq:kappadefholo}. The reader bothered by the factor of $1/T$ in the definition of $\bar{\kappa}$ could just consider the limit $T\ll\omega$ rather than strictly $T=0$, or consider the energy conductivity $\kappa_\epsilon = T\bar\kappa$ where this factor is absent and which has a regular $T=0$ limit.

We perform the following field redefinition and change of radial variable:
\begin{equation}
\bar \Pi_x=\frac{1}{\sqrt{YC^{d/2}}}\left(\frac{C^{d/2+1}}{\sqrt{BD}}\left(\frac{D}{C}\right)'\right)\Pi_x\,,\quad\quad \frac{d\bar r}{dr}=\sqrt{\frac{B}D},
\end{equation}
after which the equation for $\Pi_x$ becomes a Schrödinger equation:
\begin{equation}
\label{SeqPi}
\frac{d^2\bar \Pi_x}{d\bar r^2}+(\omega^2-V_s)\bar \Pi_x=0\,.
\end{equation}
The full expression for the Schrödinger potential $V_s$ can easily be obtained but is rather lengthy, so we do not report it.
In this form, it is easy to prove that there is a radially conserved quantity
\begin{equation}
\label{consfluxPi}
\mathcal F\equiv i\left(\bar\Pi_x^\star\partial_{\bar r}\bar\Pi_x-\bar\Pi_x\partial_{\bar r}\bar\Pi_x^{\star}\right),
\end{equation}
for real values of $\omega$. Evaluating it near the boundary, we find that $\mathcal{F}$ is related to the thermal conductivity (up to a constant prefactor) by
\begin{equation}
\label{eq:thcondflux}
\text{Re}\;\bar\kappa(\omega)=\frac{m^2}{T}\frac{\mathcal{F}}{\omega \left|\Pi_x^{(0)}\right|^2},
\end{equation}
where $\Pi_x^{(0)}$ is the value of $\Pi_x$ at the AdS boundary, which we can set to $1$ by the scaling symmetry of the perturbation equation. 

We now use the same matching argument as before. In the zero temperature IR solution that is valid in the deep interior, the Schrödinger potential is still $V_s=V_{s,0}/(4\bar r^2)$, where the value of $V_{s,o}$ depends on the class of IR solution:
\begin{equation}
\label{Vom}
\begin{split}
z\neq1\,:&\quad V_{s,0}=\frac{(d-\theta-2+2z)(d-2-\theta+4z)}{z^2},\\
z=1\,:&\quad  V_{s,0}=(d-\theta+2\Delta_m)(d+2-\theta-2\Delta_m)
\end{split}
\end{equation}
Remarkably, these are the same values as in the nonzero density case, with the same dependence on the scaling dimension of the irrelevant coupling when $z=1$. We recall that $\Delta_m=1+\kappa\lambda/2$. The next few steps are the same as before.
The ingoing solution of the Schrödinger equation \eqref{SeqPi} with this potential is
\begin{equation}
\label{eq:Hankelsolm}
\bar \Pi=\bar \Pi_0 \sqrt{\bar r} H_\nu^{(1)}\left(\omega\bar r\right),\quad 2\nu=\sqrt{1+ V_{s,0}}=\left\{\begin{array}{ll}\frac{d-2+3z-\theta}{z}\;\;&z\neq1\\d+1-\theta-2\Delta_{m}\;\;&z=1 \end{array}\right.\,.
\end{equation}
where $\bar{\Pi}_0$ is a constant and $H_\nu^{(1)}$ is the Hankel function of the first kind. Near the boundary of the IR region of the spacetime, this solution has the form
\begin{equation}
\label{eq:IRsolnexpm}
\Pi_x(r)\sim\left\{\begin{array}{ll} \Pi_0\left(r^{0}+\#\;\omega^{3+\frac{d-\theta-2}{z}}r^{d+3z-2-\theta}+\ldots\right)&z\neq1\\\Pi_0\left(r^{0}+\#\;\left(\omega r\right)^{d+1-\theta-2\Delta_m}+\ldots\right)\;\;&z=1 \end{array}\right.
\end{equation}
where $\Pi_0$ is a constant and $\#$ denote complex constants that depend on the parameters of the IR solution but not on the irrelevant coupling $m$.
 
This leads to
\begin{equation}
\text{Re}\;\bar\kappa(\omega,T=0)\sim\frac{m^2}T\left|\Pi_0\right|^2 \omega^{2\nu-1}\sim\left\{\begin{array}{ll}\frac{m^2}T\left|\Pi_0\right|^2 \omega^{2+\frac{d-\theta-2}{z}},&z\neq1\\\frac{m^2}T\left|\Pi_0\right|^2 \omega^{d-\theta-1-2\Delta_{m}}\;\;&z=1 \end{array}\right.
\end{equation}

The final step of the calculation is to fix the overall normalisation constant $\Pi_0$ such that this solution asymptotes to $1$ at the boundary of the AdS spacetime. 

To fix the constant, we will match it to a solution which is valid in this region, but is also valid near the AdS boundary. To find this second solution, we set $\omega=0$ in the equation of motion for $\Pi_x$ and solve to obtain
\begin{equation}
\Pi_x(r)=1+\pi_1A(r)\int ^r_{\text{bdy}}dr \sqrt{\frac{B}{D}}\frac{YC^{d/2}}{\left(\frac{C^{d/2+1}}{\sqrt{BD}}\left(\frac{D}{C}\right)'\right)^2},
\end{equation}
where $\pi_1$ is a constant and we have normalised the solution so that the boundary value $\Pi_x^{(0)}=1$. Expanding this in the deep interior (notice that the integrand diverges in the interior, which justifies evaluating the integral on the IR solution rather than on the full spacetime) and then matching to \eqref{eq:IRsolnexpm} gives $\Pi_0=1$. So the final result for the thermal conductivity is
\begin{equation}
\label{eq:kappaomegalowT}
\text{Re}\;\bar\kappa(\omega,T\to0)\sim\left\{\begin{array}{ll}\frac{m^2}T \omega^{2+\frac{d-\theta-2}{z}},&z\neq1\\\frac{m^2}T\omega^{d-\theta-2\Delta_{m}}\;\;&z=1 \end{array}\right.
\end{equation}

We can compare this with the dc formula
\begin{equation}
\bar\kappa=\frac{4\pi s T}{m^2 Y_h}
\end{equation}
which at low temperatures becomes
\begin{equation}
\label{eq:kappadclowT}
\begin{split}
z\neq1:&\quad \bar\kappa(\omega=0,T\to0)\sim \frac{1}{m^2}T^{1+\frac{d-2-\theta}{z}}\\
z=1:&\quad\bar\kappa(\omega=0,T\to0)\sim  \frac{1}{m^2}T^{d-\theta-1+2\Delta_m}
\end{split}
\end{equation}
Recalling that $[m]=\Delta_m=1+\kappa\lambda/2$ vanishes when $z\neq1$ (and $\Delta_m<0$ when $z=1$), the $\omega$ dependence of \eqref{eq:kappaomegalowT} matches the $T$ dependence of \eqref{eq:kappadclowT} for $z\ne1$, but not for $z=1$. Observe that the irrelevant coupling $m$ appears very differently in the two formul\ae\, as was the case at nonzero density in section \ref{section:accondnonzerodensity}.

\section{Infra-red scaling theories in the presence of irrelevant couplings \label{section:MatchScaling}}

We will now write down a scaling theory for both translation invariant and momentum relaxing QCPs. This scaling theory is consistent with the results outlined above once the dependence on the irrelevant deformation is taken into account. In the translation invariant case, this is an improvement of the scaling theory presented in \cite{Davison:2015taa}. 

\subsection{Translation-invariant case}

\subsubsection{Scaling theory}

Having assigned anomalous scaling dimensions to $s$ and $\rho$ in section \ref{subsection:IRscalingThermo}, we can now proceed to use this information to derive scaling dimensions for the incoherent response functions of the IR critical point. This constitutes a scaling theory, and it is independent of holography. While scaling theories of this type have recently been proposed as a phenomenological basis for understanding observed properties of strange metals \cite{Hartnoll:2015sea}, previous attempts to understand the IR properties of the general class of holographic models \eqref{actionEMD} using such scaling theories have run into problems \cite{Davison:2015taa}. We will describe a scaling theory (different from that in \cite{Davison:2015taa}) that is consistent with the holographic results once the dependence of observables on the dangerously irrelevant coupling $A_0$ is carefully taken into account.

In addition to the anomalous dimensions $\theta$ and $\Phi$ that we assign to the entropy and charge densities
\begin{equation}
[s]=d-\theta\,,\quad\quad\quad\quad[\rho]=d-\theta+\Phi,
\end{equation}
our scaling theory contains the additional parameter $z$. This is the dynamical critical exponent that characterises the relative scaling of space and time at the fixed point
\begin{equation}
\label{eq:spacetimescalings}
[t]=-z\,,\quad\quad\quad\quad\quad\quad\quad\quad [x]=-1.
\end{equation}
Treating temperature as an inverse timescale, we assign
\begin{equation}
[T]=z,
\end{equation}
and then using the fact that $s$ and $T$ are conjugate variables, we find that the free energy $[f]=d-\theta+z$. This can be interpreted as an effective spatial dimensionality of $d-\theta$. From this, we can assign the dimension
\begin{equation}
\label{eq:muscalingdim}
[\mu]=[f]-[\rho]=z-\Phi.
\end{equation}

We are now going to use these to calculate the dimensions of response functions of the incoherent charge. Using the definitions \eqref{RhoInc} and \eqref{HydroSources} for $\delta\rho_{\text{inc}}$ and its source $\delta s_{inc}$, we assign them the following dimensions
\begin{equation}
\label{eq:chargedims}
[\delta\rho_{\text{inc}}]=2d+z-2\theta+\Phi\,,\qquad [\delta s_{\text{inc}}]=\theta-d-\Phi.
\end{equation}
Note that these dimensions sum to $z+d-\theta$, as expected since they are thermodynamically conjugate variables. Now using the conservation equations \eqref{IncConsEq} and the spacetime dimensions (\ref{eq:spacetimescalings}), the dimensions for the incoherent current and its source are
\begin{equation}
\label{eq:currentdims}
[\delta j_{\text{inc}}]=2d-1+2z-2\theta+\Phi\,,\qquad [\delta E_{\text{inc}}]=[\partial\delta s_{\text{inc}}]=\theta-d-\Phi+1.
\end{equation}
Finally, utilising \eqref{eq:chargedims} and \eqref{eq:currentdims}, we find the scaling dimensions
\begin{equation}
\label{IncDim}
[\chi_{\text{inc}}]=3(d-\theta)+z+2\Phi\,,\qquad [\sigma_{\text{inc}}]=3(d-\theta)-2+2z+2\Phi,
\end{equation}
for the incoherent susceptibility and conductivity. Using the Einstein relation \eqref{EinsteinInc}, the dimension of the diffusivity is then
\begin{equation}
\label{IncD}
[D]=z-2.
\end{equation}
The anomalous dimensions $\theta$, $\Phi$ have dropped out from the dimension of the diffusivity $D$, as expected from the dispersion relation $\omega=-i D k^2$. 

Let us pause to clarify how our scaling theory differs from that in \cite{Davison:2015taa}. We should only expect IR scaling to apply to quantities for which UV details of the theory are unimportant. In holographic theories, dependence on UV details is manifest when quantities cannot be written in a simple way in terms of the parameters of the IR solution that is valid in the deep interior of the spacetime. One example of a UV-sensitive quantity is the electrical conductivity $\sigma$ in our theories: this is always dominated by a UV contribution $\sim i/\omega$ due to translational symmetry. Another example is the chemical potential $\mu$, which is dominated by a $T$-independent UV contribution in holographic theories at non-zero density. For this reason, and unlike in \cite{Davison:2015taa}, we did not try to construct a scaling theory which captures the properties of $\mu$ and $\sigma$, but instead worked directly with the IR quantities $\sigma_{\text{inc}}$, $\chi_{\text{inc}}$ etc.~to which scaling should apply. 

There is one subtlety: at an intermediate stage we did assign the dimension \eqref{eq:muscalingdim} to $\mu$. For the reason we have just described, this scaling dimension is not consistent with the holographic results (for example, in the $z\ne1$ cases where $T$ is the only scale, $\mu\sim T^0$ not $T^{1-\Phi/z}$), and yet the scaling dimensions \eqref{IncDim} derived from it are consistent. There are two ways to understand why this is the case. Firstly, although $\chi_{\text{inc}}$ can be written as a sum of three terms of equal scaling dimensions (equation \eqref{ChiIncSusc}), it is dominated at low temperatures by one of them: the IR $c_\rho$ contribution. That is, $\mu$ is dominated by UV contributions in such a way that its contributions to the incoherent responses are unimportant at low temperatures, and so the naive scaling dimensions we give these responses are correct. Another way to understand it is that although the incoherent source $\delta s_{\text{inc}}$ depends on $\delta\mu$, after setting the orthogonal source $\delta s_p=0$ (as we do when calculating $\chi_{\text{inc}}$), $\delta s_{\text{inc}}=-\delta T/(\rho T)$ is an IR quantity with the appropriate scaling dimension.

\subsubsection{Comparison with holographic results}

We will now show how this scaling theory is consistent with the incoherent response functions derived earlier for both kinds of translation-invariant holographic QCPs: those with $z\ne1, \Phi=\theta-d$, and those with $z=1,\Phi=(\zeta+\theta-d)/2$. Recall that the latter have an irrelevant IR coupling with dimension $[A_0]=d-\theta+\Phi$.

From the results \eqref{eq:sigmascaleholo} and \eqref{eq:chiscaleholo} for the incoherent dc conductivity and the susceptibility, it is straightforward to find that their scaling dimensions in both kinds of holographic QPCs are equal to those predicted by the scaling theory in equation \eqref{IncDim}. For the $z\ne1$ QCPs, the only dimensionful scale is $T$ and so the scaling dimensions of the incoherent dc conductivity and susceptibility capture their $T$-dependence. This is not the case for $z=1$ QCPs: while the $T$ dependence of the incoherent dc conductivity of these cases is captured by the scaling dimension, the incoherent susceptibility depends upon the (dimensionful) irrelevant coupling $A_0$ and so its total scaling dimension is made up by a combination of powers of $T$ and powers of the coupling $A_0$. This is an example of the breakdown of naive $T$-scaling (though not of the full scaling theory) due to a dangerously irrelevant coupling.

The dependence of $\chi_{\text{inc}}$ upon the irrelevant coupling is a consequence of it being proportional to $\rho^2$. In fact, we can change the dependence of $\sigma_{\text{inc}}$ and $\chi_{\text{inc}}$ upon the dangerously irrelevant coupling by changing the overall normalisation of the incoherent density \eqref{RhoInc} by powers of $\rho$. For this reason, one should not read too much into the fact that $\sigma_{\text{inc}}$ does not depend on the irrelevant coupling but $\chi_{\text{inc}}$ does. 

However, the ratio of these quantities -- the diffusivity $D$ -- is insensitive to the overall normalisation of the incoherent density. While in our holographic theories its dimension is always equal to that predicted by the scaling theory, it is only for $z\ne1$ IR solutions that this dimension corresponds to its $T$-scaling. For $z=1$ solutions, $D$ is always sensitive to the irrelevant coupling $A_0$. Thus, although $D$ and $v_B^2\tau_L$ always have the same dimension ($-1$) for $z=1$ cases, their $T$-dependences are always different since $D$ depends on the dangerously irrelevant coupling $A_0$ whereas $v_B^2\tau_L$ does not. The same phenomenon was pointed out in \cite{Blake:2017qgd} in the context of theories without translational symmetry. Using the long timescale derived in section \ref{section:accondm=0}, we have noted in section \ref{subsection:m=0Diffusivity} that it is this timescale that appears to control the diffusivity $D$ rather than $\tau_L$, through the relation \eqref{eq:Dirr}. It depends on the irrelevant coupling precisely in the same way as $D$, and so $D$ and $v_B^2 \tau_{eq}$ (with $\tau_{eq}$ given in \eqref{eq:taueqz=1TI}) now have the same $T$-dependence.

We now move to the ac conductivity $\sigma_{\text{inc}}(\omega,T=0)$ at low $\omega$. This does not depend on $T$, but does depend on the dimensionful frequency $[\omega]=-[t]=z$. Computing the scaling dimensions of $\sigma_{\text{inc}}(\omega,T=0)$ in the holographic theories from the results \eqref{eq:sigmaincholorest}, we find that they always agree with that of the scaling theory \eqref{IncDim}. 

When $z\ne1$, the $\omega$-dependence is as naively expected from the scaling dimension, as there are no other dimensionful scales. In particular, there is $\omega/T$ scaling in these cases i.e.~the $\omega$-dependence of $\sigma_{\text{inc}}(\omega,T=0)$ is the same as the $T$-dependence of $\sigma_{\text{inc}}(\omega=0,T)$. We expect that the IR conductivity $\sigma_{\text{inc}}(\omega/T)$ is a universal function that is captured by the dynamics of the near-horizon geometry. 

In contrast to this, for $z=1$ theories the ac conductivity depends on the dimensionful coupling $A_0$ and there is only consistency with the scaling theory once this dependence is carefully taken into account. Unlike for the $T$-dependence of the dc conductivity, the $\omega$-dependence in these cases is different from that naively expected from the scaling dimension. It is because of the dependence on the dangerously irrelevant coupling that these theories do not display $\omega/T$ scaling (as previously observed in the context of translation-breaking holographic theories \cite{Donos:2014uba,Gouteraux:2014hca,Donos:2014oha}) and we do not expect that the IR conductivity is a universal function in these cases.

In summary, the incoherent transport in both classes of holographic theories is consistent with our scaling theory. However, in $z=1$ cases the coupling $A_0$ is dangerously irrelevant and so the scaling theory alone cannot be used to determine how the incoherent conductivity and susceptibility depend on $\omega$ and $T$. The power of the scaling theory is significantly reduced in such cases.

\subsection{Zero density case}

The scaling dimension of the thermal conductivity can also be derived from the same scaling assumptions. Following \cite{Hartnoll:2015sea}, we find
\begin{equation}
\label{scalingpredictionkappa}
[\bar\kappa]=d-\theta+z-2\,.
\end{equation}
This is obviously consistent with the holographic results derived in section \ref{section:transportmomrel}, equations \eqref{eq:kappaomegalowT} and \eqref{eq:kappadclowT}. As for the translation-invariant case, it is crucial to appropriately account for the presence of translation-breaking irrelevant deformations in order to match the scaling prediction \eqref{scalingpredictionkappa} with the holographic results.

In the limit $T\ll\omega$, we are not aware of an argument that allows to predict how the irrelevant coupling might affect the frequency dependence of the thermal conductivity. In the opposite (dc) limit $\omega\ll T$, we can invoke well-known memory matrix arguments \cite{forster1975hydrodynamic,PhysRevB.6.1226,Hartnoll:2012rj} which predict that
\begin{equation}
\bar\kappa=\frac{\chi_{PQ}^2}{T\chi_{PP}\Gamma}+O(\Gamma^0)=\frac{\chi_{PP}}{T\Gamma}+O(\Gamma^0)
\end{equation}
where $\chi_{PQ}$ and $\chi_{PP}$ are static susceptibilities associated to the heat and momentum currents, while $\Gamma$ is the momentum relaxation rate. This approximate formula is valid when momentum relaxation is slow, $\Gamma\ll T$. In the second equality, we have further simplified it by noting that $Q= P\equiv T^{0x}$ due to the underlying relativistic symmetry of the UV CFT. The dependence on the source of the operator $\mathcal O$ breaking translations (the irrelevant coupling in our case) can then be determined by computing $\Gamma$ through
\begin{equation}
\label{GammaMM}
 \Gamma=\frac{m^2}{\chi_{PP}}\left(\lim_{\omega\to0}\frac1\omega\text{Im}G^R_{\psi\psi}(\omega,k=0)\right)_{m=0}
\end{equation}
The leading $m$ dependence sits outside the parentheses, with the prescription that the spectral weight inside should be evaluated in the translation invariant theory, and so does not depend on $m$.
This formula follows from a straightforward application of the memory matrix formalism for the translation-breaking operators discussed in the present work, see eg \cite{Hartnoll:2012rj}. We have worked out the IR dimension of the operator $\psi$ in \eqref{DimPsis}, $\Delta_{irr}=d+2-\theta-\Delta_m$. Thus we see that (remembering to Fourier transform and setting $z=1$)
\begin{equation}
\left[\frac1\omega\text{Im}G^R_{\psi\psi}(\omega,k=0)\right]=2\Delta_{irr}-1-(d+1-\theta)=d-\theta+2-2\Delta_m=d-\theta-\kappa\lambda
\end{equation}
which implies 
\begin{equation}
[\Gamma]=2\Delta_m-(d+1-\theta)+\left[\frac1\omega\text{Im}G^R_{\psi\psi}(\omega,k=0)\right]=1
\end{equation}
as expected when $z=1$. We have also used that $[\chi_{PP}]=d+1-\theta$ when $z=1$. This is obviously true for zero density, relativistic theories for which $\chi_{PP}=s T$, but can be derived more generally by recalling that $[P]=d+1-\theta$ and the definition of $\chi_{PP}$ in terms of $G^R_{PP}$. 

The result $[\Gamma]=1$ would predict $\Gamma\sim T$ if $T$ was the only dimensionful scale. However we have learned that the presence of the irrelevant coupling modifies the $T$ dependence of $\Gamma$ to
\begin{equation}
\label{GammaMMT}
\Gamma\sim T\left(\frac{m}{T^{\Delta_m}}\right)^2\,.
\end{equation}
This is because the leading $m$ dependence is captured by the overall $m^2$ factor in \eqref{GammaMM}, while the rest of the expression only depends on $T$.

The result \eqref{GammaMMT} depends on $m$ and $T$ in precisely the same way as $1/\tau_{eq}$ that we computed in \eqref{TauEqHoloM}.

In this section, we have seen a concrete case where scaling and memory matrix arguments can be combined to determine the temperature dependence both of the thermal conductivity and momentum relaxation rate, which we have checked exactly reproduces the direct holographic computation.

\acknowledgments
S.A.G. would like to thank Tom\'as Andrade, Sofia Bazakou, Nikolaos Kaplis, Alexander Krikun, Christiana Pantelidou, Alexandre Vincart-Emard and Benjamin Withers for helpful discussions on constructing finite temperature solutions numerically in the model \eqref{actionEMD}. A study of such solutions provided some inspiration for the work discussed here. B.G. would like to thank Sašo Grozdanov, Elias Kiritsis and Nick Poovuttikul for stimulating discussions.
The work of R.~A.~D.~was supported by the Gordon and Betty Moore Foundation EPiQS Initiative through Grant GBMF\#4306, the STFC Ernest Rutherford Grant ST/R004455/1 and by the STFC consolidated grant ST/P000681/1. The work of S.A.G. was supported by the Delta-Institute for Theoretical Physics (D-ITP) that is funded by the Dutch Ministry of Education, Culture and Science (OCW).  B.G. has been partially supported during this work by the Marie Curie International Outgoing Fellowship nr 624054 within the 7th European Community Framework Programme FP7/2007-2013 and by the European Research Council (ERC) under the European Union’s Horizon 2020 research and innovation programme (grant agreements No 341222 and No 758759). R.A.D and B.G. wish to thank Nordita for hospitality during the program 'Bounding Transport and Chaos in Condensed Matter and Holography'.

\appendix

\section{Equations of motion\label{app:eoms}}

The field equations for the model \eqref{actionEMD} are (repeated $I$ indices are summed over):
\begin{eqnarray}
\label{Einstein1}
&& R_{\mu\nu} + \frac{Z}{2} \, F_{\mu\rho} F^{\rho}_{\;\;\nu} -  \frac12 \partial_\mu \phi \, \partial_\nu \phi -\frac{Y}{2}\partial_\mu\psi_I\partial_\nu\psi_I+
\frac{g_{\mu\nu}}{2} \left[\frac12(\partial \phi)^2 + V -R    +  \frac{Z}{4}  F^2 +\frac{Y}{2}\left(\partial\psi_I\right)^2  \right] = 0 \, , \nn \\
\label{scalar0}
&& \frac{1}{\sqrt{-g}}\, \partial_\mu \left(\sqrt{-g}\,  \partial^\mu \phi  \right) = \frac{1}{4} \frac{\partial Z}{\partial \phi} \, F^2 
+ \frac{\partial V}{\partial \phi} + \frac{1}{2} \frac{\partial Y}{\partial \phi}\left(\partial\psi_I\right)^2  \, , \nn \\
\label{gauge0}
&& \frac{1}{\sqrt{-g}} \, \partial_\mu \left(\sqrt{-g} \, Z \, F^{\mu\nu}\right) =0  \, , \nn\\
&& \frac{1}{\sqrt{-g}} \, \partial_\mu \left(\sqrt{-g} \, Y \, \partial^\mu\psi_I\right)=0 \, .
\end{eqnarray}

We use the ansatz
\begin{equation}
\label{AnsatzBack}
\ud s^2=-D(r)\ud t^2+B(r)\ud r^2+C(r)\ud \vec{x}^2,\quad A=A(r)\ud t\,,\quad \phi=\phi(r)\,, \quad \psi_I=m\delta _{Ij}x^j,
\end{equation}
and find the Maxwell,
\begin{equation}
\label{eq:EMDMaxt}
0=\left(\frac{Z C^{d/2}A'}{\sqrt{BD}}\right)',
\end{equation}
scalar,
\begin{equation}
\label{eq:EMDScal}
0=\left(C^{d/2}\sqrt{\frac{D}{B}}\phi'\right)'+Z_{,\phi}\frac{C^{d/2}(A')^2}{2\sqrt{BD}}-\sqrt{BD}C^{d/2} V_{,\phi}-\frac{d}{2}m^2\sqrt{BD}C^{d/2-1}Y_{,\phi}\,,
\end{equation}
and Einstein equations,
\begin{equation}
\label{eq:EMDEE1}
0=\left(\frac{C^{d/2}}{\sqrt{BD}}D'\right)'+\frac{2}{d}C^{d/2}\sqrt{BD}V-\frac{\left(d-1\right)}{d}\frac{C^{d/2}Z(A')^2}{\sqrt{BD}}\,,
\end{equation}
\begin{equation}
\label{eq:EMDEE2}
0=\left(\frac{C'}{\sqrt{BCD}}\right)'+\frac{1}{d}\sqrt{\frac{C}{BD}}(\phi')^2\,,
\end{equation}
\begin{equation}
\label{eq:EMDEE3}
0=(\phi')^2-\frac{C'}{C}\left(d\frac{D'}D+\frac{d\left(d-1\right)}{2}\frac{C'}C\right)-\frac{Z(A')^2}{D}-2BV-dm^2\frac{B}{C}Y\,.
\end{equation}
The axion equations are trivially satisfied by our ansatz.

By combining these equations, we find the important identity
\begin{equation}
\frac{d}{dr}\left[\frac{C^{d/2+1}}{\sqrt{BD}}\left(\frac{D}{C}\right)'\right]=\frac{C^{d/2}Z{A'}^2}{\sqrt{BD}}+m^2\sqrt{BD}C^{d/2-1}Y,
\end{equation}
relating the metric to the matter sources. The quantity on the left hand side shows up repeatedly in the holographic transport equations we study, and we use this identity to rewrite these equations to make their dependence on the matter fields more explicit. It is by doing this that we are able to cleanly extract how the observables near $z=1$ quantum critical states specifically depend on the dangerously irrelevant couplings sourced by the matter fields. 

\bibliography{incoherent-biblio}
\end{document}